\documentclass[leqno,oneside]{article}
\usepackage{amsfonts}
\usepackage{amssymb}
\usepackage{amsmath}
\newcommand{\field}[1]{\mathbb{#1}}
\newcommand{\bn}{\mbox{\boldmath$\na$}}
\newcommand{\g}{\tilde{g}}
\newcommand{\ben}{\begin{displaymath}}
\newcommand{\een}{\end{displaymath}}
\newcommand{\Ti}{\tilde{T}}
\newcommand{\Kt}{\tilde{\hat{K}}}
\newcommand{\K}{\tilde{K}}

\newcommand{\N}{\tilde{N}}

\newcommand{\n}{\noindent}
\newcommand{\beq}{\begin{equation}}
\newcommand{\eeq}{\end{equation}}
\newcommand{\bc}{\begin{center}}
\newcommand{\ec}{\end{center}}
\newcommand{\na}{\nabla}
\newcommand{\dt}{{\bf \Pi}}
\newcommand{\W}{{\bf W}}
\newcommand{\La}{\Lambda}

\newcommand{\dod}{\,\dot{}}

\newcommand{\Nu}{\mathcal{V}}
\newcommand{\Ef}{{\mathcal{E}}_{1}}
\newcommand{\ep}{\hspace{\stretch{1}}$\Box$

\vspace{0.2cm}}
\usepackage{epsfig}
\usepackage{latexsym}
\usepackage{mathrsfs}

\newtheorem{Prop}{{\bf Proposition}}
\newtheorem{T}{{\bf Theorem}}

\newtheorem{Remark}{\bf Remark}

\newtheorem{Obs}{\it Observation}

\begin{document}

\begin{center}
{\Large\bf The ground state and the long-time evolution in the CMC Einstein flow}

\vspace{0.2cm}

{\large Martin Reiris}\footnote{e-mail: reiris@math.mit.edu.}\\

\vspace{0.1cm}

\textsc{Math. Dep. Massachusetts Institute of Technology}\\
\end{center}
\begin{center}
{\small \bf Abstract}

\vspace{0.2cm}
\begin{minipage}[c]{12cm}
\linespread{1.1}%
\selectfont
{\small Let $(g,K)(k)$ be a CMC (vacuum) Einstein flow over a compact three-manifold $\Sigma$
with non-positive Yamabe invariant $(Y(\Sigma))$. As noted by Fischer and Moncrief,
the reduced volume $\Nu(k)=(\frac{-k}{3})^{3}Vol_{g(k)}(\Sigma)$ is monotonically decreasing in the expanding direction and bounded below by $\Nu_{\inf}=(\frac{-1}{6}Y(\Sigma))^{\frac{3}{2}}$. Inspired by this fact we define the ground state of the manifold $\Sigma$ as ``the limit" of any sequence of CMC states $\{(g_{i},K_{i})\}$ satisfying: i. $k_{i}=-3$, ii. $\Nu_{i}\downarrow \Nu_{inf}$, iii. $Q_{0}((g_{i},K_{i}))\leq \Lambda$ where $Q_{0}$ is the Bel-Robinson energy and $\La$ is any arbitrary positive constant. We prove that (as a geometric state) the ground state is equivalent to the Thurston geometrization of $\Sigma$. Ground states classify naturally into three types. We provide examples for each class, including a new ground state (the Double Cusp) that we analyze in detail. Finally consider a long
time and cosmologically normalized flow $(\g,\K)(\sigma)=((\frac{-k}{3})^{2}g,(\frac{-k}{3})K)$ where $\sigma=-(\ln -k)\in [a,\infty)$. We prove that if $\tilde{\Ef}=\Ef((\g,\K))\leq \La$ (where $\Ef=Q_{0}+Q_{1}$, is the sum of the zero and first order Bel-Robinson energies) the flow $(\g,\K)(\sigma)$ persistently geometrizes the three-manifold $\Sigma$ and the geometrization is the ground state if $\Nu\downarrow \Nu_{inf}$.}
\end{minipage}
\end{center}
\begin{center}
\begin{minipage}[c]{11cm}{\small
{\center\tableofcontents}}
\end{minipage}
\end{center}
\newpage

{\center \section{Introduction}}

Consider a cosmological space-time solution ${\bf g}$ over ${\bf M}=\Sigma\times (\sigma_{0},\infty)$ where $\Sigma$ is a compact three-manifold having non-positive Yamabe invariant $Y(\Sigma)$\footnote{The Yamabe invariant (sometimes called {\it sigma constant}) is defined as the supremum of the scalar curvatures of unit volume Yamabe metrics. A Yamabe metrics is a metric minimizing the Yamabe functional in a given conformal class.}. Suppose that the foliation $\{\Sigma\times \{\sigma\}\}$ is CMC and that $\sigma$ is the logarithmic time, namely suppose that each slice $\Sigma\times \{\sigma\}$ is of constant mean curvature $k=-e^{-\sigma}$. Consider the Einstein (CMC) flow $(g,K)(\sigma)$ where $g(\sigma)$ and $K(\sigma)$ are the induced three-metric and second fundamental form over each slice $\Sigma\times \{\sigma\}$. A natural question to ask is the following. Suppose we observe the evolution of $(g,K)$ at the cosmological scale, then, is the long time fate of $(g,K)$ (at the cosmological scale) unique, and if so, how can one characterize it?. If the answer is yes, one would naturally call the limit {\it the ground state} (at the cosmological scale) as any solution would decay to it. In this article we will present partial answers to this question. We elaborate on that below.  

It is a simple but interesting fact that (with generality) one can interpret $\frac{-k}{3}$ as equal to the Hubble parameter ${\mathcal{H}}$ of the ``universe" $({\bf g},{\bf M})$ at the ``instant of time" $\Sigma\times \{\sigma(k)\}$\cite{Rei}. This cosmological interpretation of the mean curvature $k$ (or better of $\frac{-k}{3}$) motivates the terminology of various
notions that we describe in what follows. Consider a CMC slice $\Sigma\times \{\sigma\}$. At that slice the Hubble parameter 
is thus ${\mathcal{H}}=\frac{e^{-\sigma}}{3}$. For this particular value of ${\mathcal{H}}$ scale ${\bf g}$ as ${\mathcal{H}}^{2}{\bf g}$. As it is easy to see, the state $(g,K)$ over the slice $\Sigma\times \{\sigma\}$ scales to the new state $(\g,\K)=({\mathcal{H}}^{2}g,{\mathcal{H}}K)$. In this way the Hubble parameter of the new solution ${\mathcal{H}}^{2}{\bf g}$ and over the same slice will be equal to one. A state $(\g,\K)$ with ${\mathcal{H}}=1$ (or $k=-3$) will be called a {\it cosmologically normalized state}. The flow $(\g,\K)(\sigma)=({\mathcal{H}}^{2}(\sigma)g(\sigma),{\mathcal{H}}(\sigma)K(\sigma))$ will be called the {\it cosmologically normalized Einstein CMC flow}\footnote{Cosmologically normalized flows have been considered in \cite{AM} by Andersson and Moncrief. Note however that the terminology {\it Cosmologically normalized} has been introduced in \cite{Rei2}.}. Note that the volume of $\Sigma$ relative to the metric $\g$ is given by $\Nu(\sigma)={\mathcal{H}}^{3}(\sigma)Vol_{g(\sigma)}(\Sigma)$. We will call it the {\it reduced volume}. It is a crucial and central fact observed by Fischer and Moncrief \cite{FM1} that $\Nu$ is monotonically decreasing along the expanding direction and it is bounded below by the topological invariant $(-\frac{1}{6}Y(\Sigma))^{\frac{3}{2}}$. The reduced volume is a weak quantity but its relevance is greatly enhanced if we take into account at the same time the $L^{2}_{\g}$ norm of the space-time curvature ${\bf Rm}$ relative to the CMC slices, namely the Bel-Robinson energy $\tilde{Q}_{0}=Q_{0}((\g,\K))$. Our first result in (Section \ref{GSE}) will be to show  that, assuming a uniform bound in $\tilde{Q}_{0}$, the ground state of the manifold $\Sigma$ is well defined and unique. In a geometric sense the ground state is equivalent to the Thurston geometrization of $\Sigma$. Let us be more precise on the definition of ground state (under a bound in $Q_{0}$) and its characterization. By {\it ground state} we mean ``the limit" (to be described below) of any sequence of cosmologically normalized states $\{(\g_{i},\K_{i})\}$ with $Q_{0}((\g_{i},\K_{i}))\leq \La$ ($\La$ is a positive constant) and $\Nu_{i}\downarrow \Nu_{inf}$. As was proved in \cite{Rei3}, for any CMC state $(g,K)$ the $L^{2}_{g}$-norm of $Ric$ is controlled by $|k|$, $Q_{0}$ and $\Nu$ and precisely by
\ben
\|Ric\|^{2}_{L^{2}_{g}}\leq C(|k|\Nu+Q_{0}),
\een

\n where $C$ is a numeric constant. It follows that the Ricci curvature of the sequence $\{\g_{i}\}$ is uniformly bounded in $L^{2}_{\g_{i}}$. Thus \cite{A4}, one can extract a subsequence of $\{(\Sigma,\g_{i})\}$ converging in the weak $H^{2}$-topology to a (non-necessarily complete) Riemannian manifold $(\Sigma_{\infty},g_{\infty})$. We prove that the limit space $(\Sigma_{\infty},g_{\infty})$ belongs to one among three possibilities (independently of the sequence $\{(g_{i},K_{i})\}$). In general terms (see Section \ref{GS} for a more elaborate description of the ground state) the three cases are: 

\vspace{0.2cm}
\begin{enumerate}
\item (Called {\it Case $Y(\Sigma)=0$}), $\Sigma_{\infty}=\emptyset$; 

\item (Called {\it Case $Y(\Sigma)<0$ (I))}, $\Sigma_{\infty}=H$ is a hyperbolic manifold and $g_{\infty}=g_{H}$ (where $g_{H}$ is the hyperbolic metric in $\Sigma_{\infty}$); 

\item (Called {\it Case $Y(\Sigma)<0$ (II))}, $\Sigma_{\infty}=\cup_{i=1}^{i=n}H_{i}$ where $\{H_{i}\}$ is
a finite set of (non-compact) complete hyperbolic metrics of finite volume. The limit metric $g_{\infty}$ over each $H_{i}$ is equal to $g_{H,i}$ (where $g_{H,i}$ is the hyperbolic metric of $H_{i}$). The two-tori transversal to the hyperbolic cusps of each manifold $H_{i}$ embed uniquely (up to isotopy) and incompressibly (the $\pi_{1}$ injects) in $\Sigma$.  

\end{enumerate}

\n In the second and third cases $K_{i}$ converges to $-g_{H,i}$ weakly in $H^{1}$. One can also describe the notion of ground state in terms of {\it geometrizations}. This viewpoint will be fundamental in Section \ref{LTGEF}. Recall that for any Riemannian space $(\Sigma,g)$ the $\epsilon$-thick (thin) part $\Sigma^{\epsilon}$ ($\Sigma_{\epsilon}$) of $\Sigma$ is defined as the set of points $p$ in $\Sigma$ where the volume radius\footnote{Given a point $p$ in $\Sigma$ the volume radius $\nu(p)$ at $p$ is defined as the supremum of all $r>0$ such that $Vol(B(p,r))\geq \mu r^{3}$ for some fixed (but arbitrary) $\mu>0$. We define
$\underline{\nu}=inf_{p\in \Sigma}\nu(p)$ and $\overline{\nu}=sup_{p\in \Sigma}\nu(p)$. We will be using these definitions later.} $\nu(p)$ is bigger (less) or equal than $\epsilon$. Say now that $\{\g(\sigma)\}$ is a continuous ($\sigma\in [\sigma_{0},\infty)$) or discrete ($\sigma\in \{\sigma_{0},\sigma_{1},\ldots\}$)
family of Riemannian metrics on $\Sigma$. We say that $\{(\Sigma,\g(\sigma))\}$ persistently geometrizes $\Sigma$ iff there is $\epsilon(\sigma)>0$ such that $\Sigma^{\epsilon(\sigma)}_{\g(\sigma)}$ is persistently diffeomorphic to either, the empty set, or, the $\epsilon(\sigma)$-thick part of a single compact hyperbolic manifold ($(H,\g_{H})$), or, the $\epsilon(\sigma)$-thick part of a finite set of (non-compact) complete hyperbolic metrics of finite volume ($\cup_{i=1}^{i=n}(H_{i},\g_{H,i})$). The $\epsilon(\sigma)$-thin
parts $\Sigma_{\g(\sigma),\epsilon(\sigma)}$ on the other hand are persistently diffeomorphic to 
either, the empty set, or, a single graph manifold ($G$), or, a finite set of graph manifolds with toric boundaries ($\cup_{i=1}^{i=n}G_{i}$). In quantitative terms $\{\g(\sigma)\}$ geometrizes $\Sigma$ iff either
\begin{enumerate}
\item $\overline{\nu}_{\g(\sigma)}(\Sigma)\rightarrow 0$ as $\sigma$ goes to infinity (in which case there is only one persistent $G$ 
piece) or
\item $\underline{\nu}_{\g(\sigma)}(\Sigma)\geq \nu_{0}>0$ as $\sigma$ goes to infinity (in which case there is only one 
persistent $H$ piece) and there is
a continuous function $\varphi:(\sigma_{0},\infty)\times H\rightarrow \Sigma$, differentiable in the second factor, 
such that $\|\varphi^{*}\g(\sigma)-\g_{H}\|_{H^{2}_{\g_{H}}}\rightarrow 0$ as $\sigma$ goes to infinity, or
\item the volume radius collapses in some regions and remains bounded below in some others (in which case 
there are a set of $G$ pieces $G_{1},\ldots,G_{j}$ and a set of $H$ pieces $H_{1},\ldots,H_{k}$) and for any 
$\epsilon>0$ and for
any $H$ piece $(H_{i},\g_{Hi})$ there is a continuous function $\varphi_{i}:(\sigma_{0},\infty)\times 
H^{\epsilon}_{i}\rightarrow \Sigma$, differentiable in the second factor such that $\|\varphi_{i}^{*}\g(\sigma)-\g_{Hi}\|_{H^{2}_{\g_{Hi}}}
\rightarrow 0$ as $\sigma$ goes to infinity. 

\end{enumerate}

\n It is clear that cases 1,2 and 3 above correspond respectively to the three possible cases (1,2 and 3) of ground states defined before.

\begin{figure}[h]
\centering
\includegraphics[height=10cm,angle=-90]{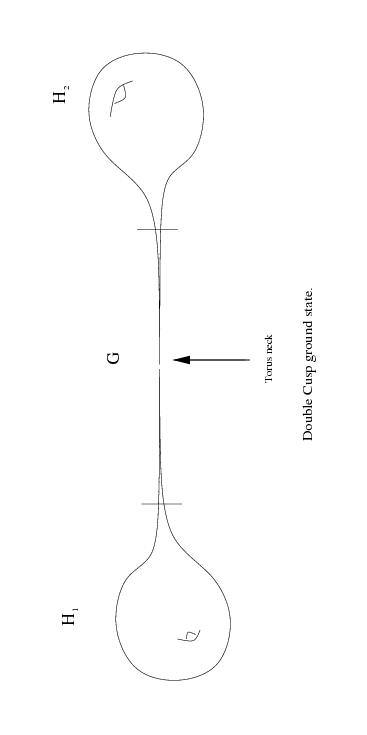}
\caption[A non pure ground state]{A schematic representation of the double cusp ground 
state.}
\label{DCFF}
\end{figure}

While it is easy to give examples of ground states of the type {\it Case $Y(\Sigma)=0$} and {\it Case $Y(\Sigma)<0$ (I)} (see Section \ref{E}) an example of the type {\it Case $Y(\Sigma)<0$ (II)} 
is more difficult to find. We dedicate Section \ref{DC} to describe a ground state of this type. The new ground state, that we shall call {\it Double Cusp}, consists of a family $\{\Sigma, (\g_{l},\K_{l})\}$ that we describe in what follows. The manifold $\Sigma$ is of the form $\Sigma=H_{1}\sharp G\sharp H_{2}$ where $H_{i}$, $i=1,2$ are (non-compact) hyperbolic manifolds with a hyperbolic cusp each\footnote{The construction can be easily generalized to include hyperbolic manifolds with any number of cusps.} and the manifold $G$ is
a so called {\it torus neck} $G=[-1,1]\times T^{2}$. The family $\{(\g_{l},\K_{l})\}$ is
parametrized by the metric ``length" $l$ of the neck. As $l\rightarrow \infty$ the geometrization
takes place. More precisely, as the length $l$ of $G$ becomes infinite, the volume radius
$\overline{\nu}(G)$ over $G$ and the total volume of $G$ collapse to zero. Over the hyperbolic sector $H_{1}$, $H_{2}$ instead, the metric $g_{l}$ converges to $g_{H_{1}}$ and $g_{H_{2}}$ respectively and in $H^{2}$. As schematic picture can be seen in Figure \ref{DCFF}.

The third part of the article (Section \ref{LTGEF}) deals with the long time evolution of the cosmologically normalized Einstein flow under the assumption that the zero and the first order Bel-Robinson energies remain uniformly bounded, namely $\tilde{\Ef}=Q_{0}((\g,\K))+Q_{1}((\g,\K))\leq \La$ for a positive constant $\La$. The main result will be to show that a long time flow $(\g,\K)(\sigma)$ with $\tilde{\Ef}\leq \La$, persistently geometrizes the manifold $\Sigma$. Moreover the geometrization is the ground state if $\Nu\downarrow \Nu_{inf}$. Using the classification of ground states (Theorem \ref{GST}) it is direct to show that ground states are stable in the following sense. For any $\La$ there is $\epsilon>0$ such that any long time cosmologically normalized flow $(\g,\K)(\sigma)$ with $\tilde{\Ef}\leq \La$ and initial data $(\g(\sigma_{0},\K(\sigma_{0}))$ with $\Nu(\sigma_{0})-\Nu_{inf}\leq \epsilon$ the flow converges to the ground state in the long time (in the sense of geometrizations). This result is not known in general if one drops the a priori (strong) assumption of a uniform bound on $\tilde{\Ef}$. However it was proved by Andersson and Moncrief \cite{AM} that if an initial data $(\g(\sigma_{0}),\K(\sigma_{0}))$ is close enough in $H^{3}\times H^{2}$ to a ground state $(H,(g_{H},-g_{H}))$ of type {\it Case $Y(\Sigma)<0$ (I)}, then $\tilde{\Ef}$ converges to zero when $\sigma\rightarrow \infty$ and $\Nu\downarrow \Nu_{inf}$, thus showing stability. We will give a proof of this fact in slightly more geometrical terms. The core of the proof is however the same.

Finally in Section \ref{HRGSGW} we present some arguments favoring the statement that
a cosmologically scaled long time CMC flow with $\tilde{\Ef}$ uniformly bounded decays necessarily to its ground state.

{\center \subsection{Background}\label{B}}

In order to avoid unnecessary repetitions we refer the reader to \cite{Rei3} for a discussion of the CMC Einstein flow as well as for the proof of several results that will be of use. The reader is also encouraged to read the original sources from which most of the background section in the reference \cite{Rei3} has been taken. 

We summarize now some basic formulae that will be used. Let us assume we have a cosmological solution\footnote{Following Bartnik a {\it cosmological solution} of the Einstein equations is a maximally globally hyperbolic solution having a compact space-like Cauchy hypersurface.} $({\bf M},{\bf g})$ with generic Cauchy hypersurface diffemorphic to $\Sigma$.
Assume $\Sigma$ is a compact three-manifold with non-positive Yamabe invariant $Y(\Sigma)$. 
Assume too that there is a CMC foliation $\Sigma\times [k_{0},0)$ inside ${\bf M}$, where 
$k$ is the mean curvature. A solution having such foliation will be called a {\it long time CMC solution}\footnote{The terminology is justified by the fact that if the manifold $\Sigma$ has non-positive Yamabe invariant then the range of $k$ (which is known to be a connected interval of the real line) cannot contain zero. If $Y(\Sigma)\leq 0$ it is conjectured that
the range of $k$ is actually $(-\infty,0)$.}. With respect to the CMC foliation the metric ${\bf g}$ splits into a space-like metric $g$, a lapse $N$ and a shift $X$. We recover the metric ${\bf g}$ from them by
\ben
{\bf g}=-(N^{2}-|X|^{2})dk^{2}+X^{*}\otimes dk+dk\otimes X^{*}+g,
\een

\n where $X^{*}=g_{ab}X^{a}$. The Einstein CMC equations in the CMC gauge (and for an arbitrary shift) are 
\begin{equation}\label{constraints1}
R=|K|^{2}-k^{2},
\end{equation}
\begin{equation}\label{constraints2}
\nabla .K=0,
\end{equation}
\begin{equation}\label{h-j1}
\dot{g}=-2NK+{\mathcal{L}}_{X}g,
\end{equation}
\begin{equation}\label{h-j2}
\dot{K}=-\nabla\nabla N+N(Ric+kK-2K\circ K)+{\mathcal{L}}_{X}K,
\end{equation}
\begin{equation}\label{lapse}
-\Delta N+|K|^{2}N=1,
\end{equation}

\n where $K$ is the second fundamental form and ${\mathcal{L}}_{X}$ is the Lie derivative operator along the vector field $X$. Sometimes we will need to use these formulas in terms of the cosmologically normalized quantities $\g={\mathcal{H}}^{2}g$, $\K={\mathcal{H}}K$ and $\tilde{N}={\mathcal{H}}^{2}N$. They will be provided without further deductions. 

The expressions for the derivative of the reduced volume with respect to logarithmic time will be central in Section \ref{LTGEF}. It is convenient to write them right away in term of cosmological normalized quantities. They are
\ben
\frac{d\Nu}{d\sigma}= -3\int_{\Sigma}1-3\tilde{N}dv_{\g}=-\int_{\Sigma}\tilde{N}|\hat{\K}|^{2}_{\g}dv_{\g}.
\een

\n The expression $\phi=3\tilde{N}-1$ is the so called {\it Newtonian potential} and it is sometimes a better quantity to work rather than the lapse $N$. 

Let us give now the basic elements of Weyl fields and Bel-Robinson energies. Again in this case the reader is encouraged to read the reference \cite{CK} for a complete account. A Weyl field is a traceless $(4,0)$ space-time 
tensor field having the symmetries of the curvature tensor ${\bf Rm}$. We will denote them by ${\bf W}_{abcd}$ or simply 
${\bf W}$. As an example, the Riemann tensor in a vacuum solution of the Einstein equations is a Weyl field that we will be denoting by 
${\bf Rm}={\bf W}_{0}$ (we will use indistinctly either ${\bf Rm}$ or $\W_{0}$). The covariant derivative of a Weyl field $\bn_{X} {\bf W}$ for an arbitrary vector field $X$
is also a Weyl field. We will be using the Weyl fields ${\bf W}_{0}={\bf Rm}$ and $\W_{1}=\bn_{T} {\bf Rm}$, where $T$ is the future pointing unit normal field to the CMC foliation. 

Given a Weyl tensor $\W$ define the current ${\bf J}$ by
\ben
\bn^{a}{\bf W}_{abcd}={\bf J}_{bcd},
\een

\n When ${\bf W}$ is the Riemann tensor in a vacuum solution of the Einstein equations the currents ${\bf J}$ is zero due to the Bianchi identities.

The $L^{2}$-norm of a Weyl field $\W$ with respect to the foliation will be introduced through the Bel-Robinson tensor which is defined by
\ben
Q_{abcd}({\bf W})={\bf W}_{alcm}{\bf W}_{b\ d}^{\ l\ m}+{\bf W}_{alcm}^{*}{\bf W^{*}}_{b\ d}^{\ l\ m}.
\een
    
\noindent The Bel-Robinson tensor is symmetric and traceless in all pair of index and for any pair of time-like vectors $T_{1}$ and $T_{2}$,
the quantity $Q(T_{1}T_{1}T_{2}T_{2})$ is positive (provided ${\bf W}\neq 0$). 

The electric and magnetic components of ${\bf W}$ are defined as
\beq\label{EB1}
E_{ab}={\bf W}_{acbd}T^{c}T^{d},
\eeq
\beq\label{EB2}
B_{ab}=^{*}{\bf W}_{acbd}T^{c}T^{d},
\eeq

\noindent where the left dual of $\W$ is defined by $^{*}{\bf W}_{abcd}=\frac{1}{2}\epsilon_{ablm}{\bf W}^{lm}_{\ \ cd}$. $E$ and $B$ are symmetric, traceless and null in the $T$ direction. It is also the case that $\W$ can be reconstructed from them 
(see \cite{CK}, page 143). If ${\bf W}$ is the Riemann tensor in a vacuum solution we have
\begin{equation}\label{de}
E_{ab}=Ric_{ab}+kK_{ab}-K_{a}^{\ c}K^{c}_{\ b},
\end{equation}
\begin{equation}\label{de2}
\epsilon_{ab}^{\ \ l}B_{lc}=\nabla_{a}K_{bc}-\nabla_{b}K_{ac}.
\end{equation}

\n The components of a Weyl field with respect to the CMC foliation are given by ($i,j,k,l$ are spatial indices) 
\beq\label{EB3}
{\bf W}_{ijkT}=-\epsilon_{ij}^{\ \ m}B_{mk},\ ^{*}{\bf W}_{ijkT}=\epsilon_{ij}^{\ \ m}E_{mk},
\eeq
\beq\label{EB4}
{\bf W}_{ijkl}=\epsilon_{ijm}\epsilon_{kln}E^{mn},\ ^{*}{\bf W}_{ijkl}=\epsilon_{ijm}\epsilon_{kln}B^{mn}.
\eeq

\noindent We also have 
\ben
Q(TTTT)=|E|^{2}+|B|^{2},
\een 
\ben
Q_{iTTT}=2(E\wedge B)_{i},
\een
\ben
Q_{ijTT}=-(E\times E)_{ij}-(B\times B)_{ij}+\frac{1}{3}(|E|^{2}+|B|^{2})g_{ij}.
\een

\n The operations $\times$ and $\wedge$ are provided explicitly later. The divergence of the Bel-Robinson tensor is
\ben
\begin{split}
\bn^{a}Q({\bf W})_{abcd}=&{\bf W}_{b\ d}^{\ m\ n}{\bf J}({\bf W})_{mcn}+{\bf W}_{b\ c}^{\ m\ n}{\bf J}({\bf W})_{mdn}\\
&+^{*}{\bf W}_{b\ d}^{m\ n}{\bf J}^{*}({\bf W})_{mcn}+^{*}{\bf W}_{b\ c}^{m\ n}{\bf J}^{*}(W)_{mcn}.
\end{split}
\een

\n where ${\bf J}^{*}_{bcd}=\bn^{a}(^{*}\W_{abcd})$. We have therefore
\ben
\bn^{\alpha}Q({\bf W})_{\alpha TTT}=2E^{ij}({\bf W}){\bf J}({\bf W})_{iTj}+2B^{ij}{\bf J}^{*}({\bf W})_{iTj}.
\een

\noindent From that we get the {\it Gauss equation} 
\beq\label{Gausseq}
\dot{Q}(\W)=-\int_{\Sigma}2NE^{ij}({\bf W}){\bf J}({\bf W})_{iTj}+2NB^{ij}(\W){\bf J}^{*}({\bf W})_{iTj}
+3NQ_{abTT}\dt^{ab}dv_{g}.
\eeq

\noindent $\dt_{ab}=\bn_{a}T_{b}$ is the {\it deformation tensor} and plays a fundamental role in the space-time 
tensor algebra. Its components are
\ben
\dt_{ij}=-K_{ij}, \ \ \ \dt_{iT}=0,
\een
\ben
\dt_{Ti}=\frac{\nabla_{i}N}{N}, \ \ \ \dt_{TT}=0.
\een

\n Finally we have
\begin{equation}\label{eq1}
div E({\bf W})_{a}=(K\wedge B({\bf W}))_{a}+{\bf J}_{TaT}({\bf W}),
\end{equation}
\begin{equation}\label{eq2}
div B({\bf W})_{a}=-(K\wedge E({\bf W}))_{a}+{\bf J}^{*}_{TaT}({\bf W}),
\end{equation}
\begin{equation}\label{eq3}
curl B_{ab}(\W)=E(\bn_{T}{\bf W})_{ab}+\frac{3}{2}(E(\W)\times K)_{ab}-\frac{1}{2}kE_{ab}(\W)+{\bf J}_{aTb}(\W),
\end{equation}
\begin{equation}\label{eq4}
curl E_{ab}(\W)=B(\bn_{T}{\bf W})_{ab}+\frac{3}{2}(B(\W)\times K)_{ab}-\frac{1}{2}kB_{ab}(\W)+{\bf J}^{*}_{aTb}(\W).
\end{equation}

\noindent The operations $\wedge,\ \times$ and the operators $Div$ and $Curl$ are defined through
\ben
(A\times B)_{ab}=\epsilon_{a}^{\ cd}\epsilon_{b}^{\ ef}A_{ce}B_{df}+\frac{1}{3}(A\circ B)g_{ab}-\frac{1}{3}(tr A)(tr B)g_{ab},
\een
\ben
(A\wedge B)_{a}=\epsilon_{a}^{\ bc}A_{b}^{\ d}B_{dc},
\een
\ben
(div\ A)_{a}=\nabla_{b}A^{b}_{\ a},
\een
\ben
(curl\ A)_{ab}=\frac{1}{2}(\epsilon_{a}^{\ lm}\nabla_{l}A_{mb}+\epsilon_{b}^{\ lm}\nabla_{l}A_{ma}).
\een

In what follows we describe the main results that will be used from the theory of convergence-collapse of Riemannian manifolds under $L^{2}$-bounds on the Ricci curvature.
The reader can consult the original sources \cite{A4}, \cite{Y1} and \cite{Y2}.

\vspace{0.2cm}
\begin{T}\label{funcc} Let $\{(\Sigma,g_{i})\}$ be a sequence of compact Riemannian manifolds with 
\ben
\|Ric\|_{L^{2}_{g_{i}}}+Vol_{g_{i}}(\Sigma)\leq \La,
\een

\n where $\La$ is a positive constant. Then one can
extract a sub-sequence (to be denoted also by $\{(\Sigma,g_{i})\}$) with one of the following
behaviors.

(1) ({\it Collapse}). $\overline{\nu}_{i}\rightarrow 0$ and the sub-sequence $g_{i}$ collapses along a sequence of F-structures. The manifold $\Sigma$ is in this case a graph manifold.

(2) ({\it Convergence}). $\underline{\nu}_{i}\geq \underline{\nu}_{0}>0$ and $\{(\Sigma,\{g_{i}\})\}$ converges weakly in $H^{2}$ to a $H^{2}$ Riemannian manifold $(\Sigma_{\infty}(=\Sigma), g_{\infty})$. 

(3) ({\it Convergence-Collapse}). $\underline{\nu}_{i}\rightarrow 0$ and $\overline{\nu}_{i}\geq \overline{\nu}_{0}>0$ and $\{(\Sigma,g_{i})\}$ converges weakly in $H^{2}$ to a (at most) countable union $\cup_{\alpha}(\Sigma_{\infty,\alpha},g_{\infty,\alpha})$ of $H^{2}$ (non necessarily complete) Riemannian manifolds. Moreover, for a given $\epsilon$ (sufficiently small) the manifolds $\Sigma_{g_{i},\epsilon}$ are graph manifolds with toric boundaries. 
The Riemannian-manifolds $\{(\Sigma^{\epsilon}_{g_{i}},g_{i})\}$ converge weakly in $H^{2}$ to $\cup_{\alpha} (\Sigma_{\infty,\alpha}^{\epsilon},g_{\infty,\alpha})$ (which has only a finite number of components).
 
\end{T}

\n The notion of convergence that we have assumed in the statement of the theorem is the following:
we say that $\{(\Sigma,g_{i})\}$ converges weakly in $H^{2}$ to a limit Riemannian manifold $(\Sigma_{\infty},g_{\infty})$ (as above) if for every $\epsilon>0$ there are ($H^{3}$)-diffemorphisms $\varphi_{i}:\Sigma^{\epsilon}_{\infty,g_{\infty}}\rightarrow \Sigma^{\epsilon}_{g_{i}}$ such that $\varphi_{i}^{*}g_{i}$ converges to $g_{\infty}$ in the weak $H^{2}$-topology induced by the metric $g_{\infty}$ over the space of $H^{2}$ (2,0)-tensors (over $\Sigma_{\infty}$).

We will use the notation $H^{*}_{\star}$ to denote the $*$-Sobolev space defined with respect to the structure $\star$. For instance $H^{2}_{\{x\}}$ is the 2-Sobolev space
defined with respect to a chart $\{x\}$. $H^{1}_{g}$ instead is the $1$-Sobolev space defined
with respect to the metric $g$. For more details on the notation see \cite{Rei3}.

{\center \section{The ground state and examples}\label{GSE}}

{\center \subsection{The ground state}\label{GS}}

\begin{T}\label{GST}{\rm (The ground state)}
Let $\Sigma$ be a compact three-manifold with $Y(\Sigma)\leq 0$. Say $\{(g_{i},K_{i})\}$ is a sequence of states
satisfying
\ben
\begin{array}{rl}
(1)& k_{i}=-3;\\
(2)& \Nu_{i}\downarrow \Nu_{inf}=(-\frac{1}{6}Y(\Sigma))^{\frac{3}{2}};\\
(3)&Q_{0}((g_{i},K_{i}))\leq \Lambda,
\end{array}
\een

\n where $\Lambda$ is a fixed constant. Then, there is a sub-sequence of $\{(g_{i},K_{i})\}$ 
(to be denoted also by $\{(g_{i},K_{i})\}$) for which one and only one of the following three phenomena occurs.

\vspace{0.2cm}
\n {\it Case $Y(\Sigma)=0$}. 
\begin{enumerate}
\item $\Sigma=G$ is a graph manifold. 
\item $\overline{\nu}\rightarrow 0$ and the Riemannian spaces $(\Sigma,g_{i})$ collapse with bounded $L^{2}$ curvature, along a sequence of F-structures.
\item $\Nu_{i}\downarrow \Nu_{inf}=0$
\end{enumerate}

\n {\it Case $Y(\Sigma)<0$} (I).
\begin{enumerate}
\item $\Sigma=H$ is a compact hyperbolic manifold (denote its hyperbolic metric by $g_{H}$).
\item $(\Sigma,g_{i})\rightarrow (\Sigma,g_{H})$ in the weak $H^{2}$-topology. 
\item $\Nu_{i}\downarrow Vol_{g_{H}}=(-\frac{1}{6}Y(\Sigma))^{\frac{3}{2}}$.
\end{enumerate}

\n {\it Case $Y(\Sigma)<0$} (II).
\begin{enumerate}
\item There is a set of incompressible two-tori $\{T^{2}_{i},i=1,\ldots i_{T}\}$ embedded in $\Sigma$
and cutting it into a set $\{H_{i},i=1,\ldots i_{H}\}$ of manifolds admitting a 
complete hyperbolic metric of finite volume 
(in its interior) and a set $\{G_{i},i=1,\ldots i_{G}\}$ of graph manifolds. The tori $T^{2}_{i}$ are unique
up to isotopy.
\item $(\Sigma,g_{i})\rightarrow \cup_{i=1}^{i=i_{H}}(H_{i},g_{H,i})$ in the weak $H^{2}$-topology.
\item $\Nu_{i}\downarrow \sum_{i=1}^{i=i_{H}}Vol_{g_{H,i}}(H_{i})=(-\frac{1}{6}Y(\Sigma))^{\frac{3}{2}}$.
\end{enumerate}

In each of the three cases above the norms $\|Ric_{g_{i}}\|_{L^{2}_{g_{i}}},\ \|K_{i}\|_{H^{1}_{g_{i}}}$, and $\|K_{i}\|_{L^{4}_{g_{i}}}$ remain uniformly bounded and the norms $\|\hat{K}_{i}\|_{L^{2}_{g_{i}}}$, $\|R_{g_{i}}+6\|_{L^{1}_{g_{i}}}$ converge to zero. Moreover in the regions of convergence (the hyperbolic sector in (I) and (II)) the scalar curvature $R_{g_{i}}$ converges in the strong $L^{2}$-topology to $-6$.

Finally, two different sub-sequences of the original sequence $\{(g_{i},K_{i})\}$ as above have the same 
behavior.

\end{T}

\n {\bf Proof:} 

Recall the following inequalities in \cite{Rei3}. From Proposition 6 in \cite{Rei3} we have
\beq\label{F1}
\int_{\Sigma}2|\na \hat{K}|^{2}+|\hat{K}|^{4}dv_{g}\leq C(|k|({\mathcal{V}}-{\mathcal{V}}_{inf})+Q_{0}),
\eeq

\n and from Proposition 7
\beq\label{F2}
\int_{\Sigma}|k|^{2}|\hat{K}|^{2}dv_{g}\leq C(|k|({\mathcal{V}}-{\mathcal{V}}_{inf})+
(|k|({\mathcal{V}}-{\mathcal{V}}_{inf})Q_{0})^{\frac{1}{2}}).
\eeq

\n This in particular implies the inequality
\beq\label{R}
\int_{\Sigma}|k|^{2}(R_{g}+\frac{2}{3}k^{2})dv_{g}\leq C(|k|({\mathcal{V}}-{\mathcal{V}}_{inf})+
(|k|({\mathcal{V}}-{\mathcal{V}}_{inf})Q_{0})^{\frac{1}{2}}).
\eeq

\n From Proposition 8 we have
\beq\label{Ric}
\|\hat{Ric}\|_{L^{2}_{g}}\leq C((|k|({\mathcal{V}}-{\mathcal{V}}_{inf})+
(|k|({\mathcal{V}}-{\mathcal{V}}_{inf})Q_{0})^{\frac{1}{2}}+Q_{0}).
\eeq

\n and from Proposition 9 also in \cite{Rei3}
\beq\label{F3}
\int_{\Sigma}|\nabla R|^{\frac{4}{3}}+R^{2}dv_{g}\leq C(|k|{\mathcal{V}}+Q_{0}).
\eeq

\n Recall when using the formulas above that we will be dealing with a sequence $\{(\g_{i},\K_{i})\}$ with $k_{i}=-3$.

{\it Case $Y(\Sigma)=0$}. From (\ref{Ric}) we see that the $L^{2}_{g_{i}}$ norm of $Ric_{g_{i}}$ remains
uniformly bounded. This case then follows from Theorem \ref{funcc}.

{\it Case $Y(\Sigma)<0$}. First we note that there must be a constant $\overline{\nu}_{0}>0$ such that 
$\overline{\nu}_{i}\geq \overline{\nu}_{0}$ otherwise one can extract a sub-sequence of $\{g_{i}\}$ which collapses with bounded volume and curvature. Theorem \ref{funcc} then implies that $\Sigma$ is a graph manifold and therefore of zero Yamabe invariant which is a contradiction. Cases (I) and (II) will be
distinguished according to whether there is a sub-sequence of $\{g_{i}\}$ with $\underline{\nu}_{i}\rightarrow 0$ or not. We do that below.

(I). Suppose there exists $\underline{\nu}_{0}>0$ such that $\underline{\nu}_{i}\geq \underline{\nu}_{0}$. Then by Theorem \ref{funcc} we can extract a sub-sequence of $\{g_{i}\}$ converging in the weak $H^{2}$-topology to a compact Riemannian manifold $(\Sigma,g_{\infty})$. From (\ref{R}) we deduce that $R_{g_{\infty}}=-6$. Let us see that $g_{\infty}$ is hyperbolic. Note that $\int_{\Sigma}R_{g_{\infty}}^{2}dv_{g_{\infty}}=|Y(\Sigma)|^{2}$. Consider the quadratic functional ${\mathcal{R}}$ from $H^{2}$-metrics into the reals given by  
\ben
g\rightarrow Vol_{g}(\Sigma)^{\frac{1}{3}}\int_{\Sigma}R_{g}^{2}dv_{g}.
\een

\n It is known \cite{ASCS} that if $Y(\Sigma)<0$ the infimum of ${\mathcal{R}}$ is given by $|Y(\Sigma)|^{2}$. Thus it must be $\delta {\mathcal{R}}|_{g_{\infty}}=0$. Let us compute the variation of ${\mathcal{R}}$ at $g=g_{\infty}$ and for variations which preserve the local volume. Consider then an arbitrary path of metrics $g(\lambda)$ with $g(\lambda=0)=g_{\infty}$ and $(dv_{g(\lambda)})'=0$ (and Frechet derivative $g'=h$ in $H^{2}$). From $(dv_{g})'=0$ we get $tr_{g}h=0$. Recall the variation of the scalar curvature
\ben
\delta_{h}R_{g}=\Delta tr_{g}h+\delta\delta h-<Ric,h>.
\een

\n From it we get
\ben
\delta_{h}{\mathcal{R}}_{g}|_{g=g_{\infty}}=-Vol_{g_{\infty}}^{\frac{1}{3}}2R_{\infty}\int_{\Sigma}<\hat{Ric}_{g_{\infty}},h>dv_{g_{\infty}}.
\een

\n Thus $\hat{Ric}_{g_{\infty}}=0$ and $g_{\infty}$ is hyperbolic. Therefore this case corresponds to {\it Case $Y(\Sigma)<0$} (I). 

(II). Suppose $\limsup \underline{\nu}_{i}=0$. Consider a $H^{2}$-weak limit of $(\Sigma,g_{i})$. Denote it
by $(\Sigma_{\infty},g_{\infty})$. Recall that $\Sigma_{\infty}$ may have infinitely many connected components and that $g_{\infty}$ may not be complete on them. Note that $\Sigma_{\infty}$ is non-empty
as we have $\overline{\nu}_{i}\geq \overline{\nu}_{0}>0$. For every $i$ consider the metric $g_{Y,i}$ in the conformal class of $g_{i}$ with scalar curvature $R_{Y}=-6$. Writing $g_{Y,i}=\phi_{i}^{4}g_{i}$, the conformal factor $\phi_{i}$ satisfies the equation
\beq\label{Yam}
R_{Y}\phi_{i}^{5}=-8\Delta_{g_{i}}\phi_{i}+R_{g_{i}}\phi_{i}.
\eeq

\n From the maximum principle we get $\phi_{i}\leq 1$. Thus
\ben
0\leq Vol_{g_{i}}(\Sigma)-Vol_{g_{Y,i}}(\Sigma)\leq Vol_{g_{i}}(\Sigma)-Vol_{inf}.
\een

\n It follows from the fact that $\Nu_{i}\downarrow \Nu_{inf}$ that 
\ben
0\leq \int_{\Sigma}1-\phi_{i}^{6}dv_{g_{i}}\rightarrow 0.
\een

\n and in particular $\int_{\Sigma}(1-\phi_{i})^{6}dv_{g_{i}}\rightarrow 0$. Note that 
\ben
Vol_{g_{Y,i}}^{\frac{1}{3}}\int_{\Sigma}R_{g_{Y,i}}^{2}dv_{g_{Y,i}}\rightarrow |Y(\Sigma)|^{2}.
\een

\n We will exploit this fact in what follows. Pick an arbitrary point $p\in\Sigma_{\infty}$. We will show that $\hat{Ric}_{g_{\infty}}|_{B(p,\nu_{g_{\infty}}(p)/2)}=0$. As the point $p$ is arbitrary this would show that $\hat{Ric}_{g_{\infty}}=0$ and thus $g_{\infty}$ is hyperbolic. First note that by (\ref{R}) it is $R_{g_{\infty}}=R_{Y}=-6$. Also by (\ref{F3}) and the compact embedding $H^{1,4/3}\hookrightarrow L^{2}$ we see that $R_{g_{i}}\rightarrow R_{Y}$ strongly in $L^{2}$ on compact sets of $\Sigma_{\infty}$. Pick a sequence $\{p_{i}\}$ of points $p_{i}\in \Sigma$ such that $(B_{g_{i}}(p_{i},\nu_{g_{i}}(p_{i})),p_{i},g_{i})$ converges to $(B_{g_{\infty}}(p,\nu_{g_{\infty}}(p)),p,g_{\infty})$. Let us write equation (\ref{Yam}) in the form
\beq\label{Yam2}
8\Delta_{g_{i}}\phi_{i}=R_{g_{i}}\phi_{i}-R_{Y}\phi_{i}^{5},
\eeq
 
\n and prove that the right hand side of it is converges to zero in $L^{2}$ and over the sequence of balls $B_{g_{i}}(p_{i},\nu_{g_{i}}(p_{i}))$ (denote them by $B_{i}$). Write 
\beq\label{L20}
\begin{split}
\int_{B_{i}}|R_{Y}\phi_{i}^{5}-R_{g_{i}}\phi_{i}|^{2}dv_{g_{i}}\leq &\int_{B_{i}}|R_{Y}\phi_{i}^{4}-R_{g_{i}}|^{2}dv_{g_{i}}\\
&\leq\int_{B_{i}}2|R_{Y}|^{2}|\phi^{4}-1|^{2}+2|R_{Y}-R_{g_{i}}|^{2}dv_{g_{i}}.
\end{split}
\eeq 

\n We have
\ben
\int_{B_{i}}|\phi_{i}^{4}-1|^{2}dv_{g_{i}}\leq \int_{B_{i}}|\phi_{i}^{4}-1|dv_{g_{i}}\rightarrow 0,
\een

\n From this and the fact that $R_{g_{i}}$ converges to $R_{Y}$ strongly in $L^{2}$ over $B_{i}$ we have that the right hand side of equation (\ref{L20}) goes to zero as claimed. By elliptic regularity $\|\phi_{i}-1\|_{H^{2}_{g_{i}}(B_{g_{i}}(p_{i},\frac{2}{3}\nu_{g_{i}}(p_{i}))}$ converges to zero. As a result $(B_{g_{i}}(p_{i},\frac{2}{3}\nu_{g_{i}}(p_{i}),p_{i},g_{Y,i})$ converges weakly in $H^{2}$ to $(B_{g_{\infty}}(p,\frac{2}{3}\nu_{g_{\infty}}(p)),p,g_{\infty})$. As a consequence we have that 
\ben
\int_{\Sigma}<\hat{Ric}_{g_{Y,i}},h_{i}>_{g_{Y,i}}dv_{g_{Y,i}}\rightarrow \int_{\Sigma_{\infty}}<Ric_{g_{\infty}},h>_{g_{\infty}}dv_{g_{\infty}},
\een

\n for any traceless tensor $h$ (in $H^{2}$) with support in $B_{g_{\infty}}(p,\nu_{g_{\infty}}(p)/2)$ and traceless tensors $h_{i}$ (in $H^{2}$) with support in $B_{g_{i}}(p_{i},\nu_{g_{i}}(p_{i})/2)$ converging strongly in $H^{2}$ to $h$. Thus $\delta_{h_{i}}{\mathcal{R}}|_{g=g_{Y,i}}\rightarrow \delta_{h}{\mathcal{R}}|_{g=g_{\infty}}$. Therefore if $\hat{Ric}_{g_{\infty}}\neq 0$ in $B_{g_{\infty}}(p,\nu_{g_{\infty}}(p)/2)$ we can lower the infimum of $|Y(\Sigma)|^{2}$ for the functional ${\mathcal{R}}$ over the three-manifold 
$\Sigma$. 

We prove now that $g_{\infty}$ is complete. Let $s$ be an incomplete geodesic in $\Sigma_{\infty}$. Fix $p\in s$. Let $S_{2}$ be a transversal geodesic-two-simplex in $\Sigma_{\infty}$ and having $p$ in its interior. For $q\in s$ (in the incomplete direction and close to $p$) consider the three-simplex 
$S_{3}(q)$ formed by all geodesics joining $q$ with a point in $S_{2}$. Observe that because $(\Omega, g_{\infty})$ is hyperbolic and $s$ 
has finite length (in the incomplete direction) every $r\in\partial S_{3}(x)$ has a cone $C_{3}(r)$ inside and of size bounded 
below\footnote{Given a point $x$ in a Riemannian manifold $(\Sigma, g)$ a cone of size $(\alpha, l)$ ($l<inj_{x}g$) in $\Sigma$ is the 
image under the exponential map of a cone of size $(\alpha, l)$ (segments from $x$ in $T_{x}\Sigma$ having length $l$ and forming an 
angle $\alpha$ with a given segment)}. Now as $q$ approaches the end of $s$, we can find a sequence of points $q_{i}$ and $r(q_{i})\in \partial S_{3}(q_{i})$ with 
$\nu_{g_{\infty}}(r(q_{i}))\rightarrow 0$ and having a cone inside $S_{3}(r(q_{i}))$ of size bounded below. The blow up limit of the pointed space 
$(\Sigma_{\infty},r(q_{i}),\frac{1}{\nu(r(q_{i}))^{2}}g_{\infty})$ has $\nu(x)=1$ and is complete, flat and having a cone of size 
$(\alpha, \infty)$ inside. It must be $\field{R}^{3}$ which is a contradiction.  

We can conclude then that $\Sigma_{\infty}$ consist of a finite number of connected components $H_{i},i=1,\ldots,i_{H}$ each one admitting a complete hyperbolic metric of finite
volume $g_{H,i}=g_{\infty}$. Observe that it must be $Vol_{g_{\infty}}(\Sigma_{\infty})=Vol_{\inf}$ (and not strictly less than $Vol_{inf}$) for otherwise (see \cite{CHG}) one can find a sequence of metrics $\tilde{\tilde{g}}_{i}$ in $\Sigma$ and with bounded $L^{\infty}_{\tilde{\tilde{g}}_{i}}$ curvature, converging to $\cup_{i=1}^{i=i_{H}}(H_{i},g_{H,i})$ and with $Vol_{\tilde{g}_{i}}(\Sigma)\rightarrow Vol_{g_{\infty}}(\Sigma_{\infty})$ and thus lowering the value $|Y(\Sigma)|^{2}$ for the infimum of ${\mathcal{R}}$.

Now, pick a transversal torus for each one of all the hyperbolic cusps of the Riemannian manifolds $(H_{i},g_{H,i})$. Denote them by $\{T_{i},i=1,\ldots,i=i_{T}\}$. Each one of the tori $T_{i}$ can be embedded (up to isotopy) inside $\Sigma$. As proved in \cite{A3} (Theorem 2.9) if one of the tori is compressible one can again lower the infimum value for ${\mathcal{R}}$. Thus the tori $T_{i}$ are all incompressible. As shown in \cite{A3} (page 156) the set of tori $\{T_{i},i=1,\ldots,i_{T}\}$ (of a strong geometrization as this) is unique up to isotopy.      

The rest of the claims in the Theorem follow from equations (\ref{F1})-(\ref{F3}).\ep

{\center \subsection{Examples}\label{E}}

Examples of ground states (namely sequences $\{(g_{i},K_{i})$ of cosmologically normalized
states with $\Nu_{i}\downarrow \Nu_{inf}$ and $Q_{0}\leq \Lambda$) and of the types 
{\it Case $Y(\Sigma)=0$} or {\it Case $Y(\Sigma)<0$ (I)} (in Theorem \ref{GS}) are easy to find. We will show that soon below. An example of a ground state of the type {\it Case $Y(\Sigma)<0$ (II)} is more difficult to find and will be discussed in a separate section (the next Section \ref{DC}).  

{\it Case $Y(\Sigma)=0$}. Take any two-surface $\Sigma_{gen}$ of genus greater or equal than one. Consider the three-manifold $\Sigma=\Sigma_{gen}\times S^{1}$. Denote by $l^{2}ds^{2}$ the metric on $S^{1}$ with total length $l$ and denote by $g_{gen}$ a metric on $\Sigma_{gen}$ of scalar curvature $-6$. An example of a ground state of the type {\it Case $Y(\Sigma)=0$} is given by the sequence of states $\{(g_{l},-g_{l})\}$ on $\Sigma$ where $g_{l}=g_{gen}\times l^{2}ds^{2}$ and $l\rightarrow 0$. 

{\it Case $Y(\Sigma)<0$ (I)}. Take any compact hyperbolic manifold $\Sigma$ with hyperbolic metric $g_{H}$. The constant sequence of states $\{(g_{H},-g_{H})\}$ is an example of a ground state of type {\it Case $Y(\Sigma)<0$ (I)}.  

\vspace{0.2cm}
{\center \subsection{The double cusp}\label{DC}} 
  
Say $(H_{1},g_{H_{1}})$ and $(H_{2},g_{H_{2}})$ are two complete hyperbolic metrics of finite volume and suppose that each one has, for the sake of concreteness, only one
hyperbolic cusp. Denote the cusps as $C_{1}$ and $C_{2}$. Denote by $(g_{H_{i}},-g_{H_{i}})$
the flat cone states on $H_{i}$, $i=1,2$. Recall that the metrics $g_{H_{i}}$ on the cusps are of the form $g_{H_{i}}=dx^{2}+e^{2x}g_{T,i}$ where $g_{T_{i}}$ is a flat (and $x$-independent) metric on the tori $T^{2}$ transversal to the cusps $(-\infty,a]\times T^{2}$. Consider now a torus-neck, namely the manifold $G=[-l,l]\times T^{2}$ with a $T^{2}$-invariant metric $g_{G}$. For any $x_{0}<a$ we will find an state $(g_{G},K_{G})$ on $G$ which, at the boundary
$\partial [-l,l]\times T^{2}=\{-l\}\times T^{2} \cup \{l\}\times T^{2}$, approximates
to any given desired order the flat cone states of $H_{1}$ and $H_{2}$ at $x=x_{0}$.
Once this is done we will glue $(g_{H_{1}},-g_{H_{1}})$, $(g_{G},K_{G})$ and $(g_{H_{2}},-g_{H_{2}})$ to get an state over $H_{1}\sharp G\sharp H_{2}$ (satisfying the constraints equations). As $x_{0}\rightarrow -\infty$ these ``double cusp" states display the behavior of a ground state of type {\it Case $Y(\Sigma)<0$ (II)}. A schematic picture can be seen in Figure \ref{DCFF}. Note that the states $(g_{G},K_{G})$, being $T^{2}$-symmetric, are Gowdy and therefore explicitly tractable.

The construction is organized as follows. In Section \ref{TNPOL} we find a (Gowdy) polarized space-time solution on $\field{R}\times \field{R}\times T^{2}$. Once this is done, we find
in Section \ref{CCP} a foliation of $\field{R}\times \field{R}\times T^{2}$ whose states display (when suitable normalized) a convergence-collapse behavior of the type {\it Case $Y(\Sigma)<0$ (II)}. Although the states found in this foliation are not CMC, we will see in Section \ref{GL} that it is possible to find a CMC foliation whose CMC states are not far from those found before and displaying the same convergence-collapse behavior. In Section \ref{GTNNP} we find (Gowdy) non-polarized space-time solutions on $\field{R}\times \field{R}\times T^{2}$. One can then repeat the analysis done in Sections \ref{CCP} and \ref{GL} to find, for each space-time non-polarized solution, a CMC foliation displaying a convergence-collapse picture of the type {\it Case $Y(\Sigma)<0$ (II)}. The family of polarized states that we will construct is sufficient to join two arbitrary flat cone cusp sates $(C_{1},(g_{H_{1}},-g_{H_{1}}))$ and $(C_{2},(g_{H_{2}},-g_{H_{2}}))$. Suppose now we have two flat cone states $(H_{i},(g_{H_{i}},-g_{H_{2}}))$ 
having a hyperbolic cusp each that we want to join through a state in a torus-neck. Having fixed $x_{0}$ and a given error $\epsilon$, suppose we have found a state (polarized or not) $(g_{G},K_{G})$ in a torus-neck $G$, which is compatible (up to the error $\epsilon$) at its ends with the flat cone cusps $(C_{1},(g_{H_{1}},-g_{H_{1}}))$ and $(C_{2},(g_{H_{2}},-g_{H_{2}}))$ at $x=x_{0}$. We will perform the gluing of $(H_{1},(g_{H_{1}},-g_{H_{1}}))$, $(g_{G},K_{G})$ and $(H_{2},(g_{H_{2}},-g_{H_{2}}))$ as follows. First we glue (keeping the $T^{2}$-symmetry) the metrics $g_{H_{i}}$, $i=1,2$ and $g_{G}$ on an interval ($[a,b]\times T^{2}$) of length one in each one of the necks and centered at $x=x_{0}$. Denote the new metric by $g_{\sharp}$. Then we find a transverse traceless tensor $\hat{K}_{TT}$ with respect to $g_{\sharp}$ and equal to $-g_{H_{i}}$ or $K_{G}$
outside the intervals where the metrics were glued. Using the data $(g_{\sharp},\hat{K}_{TT})$ we appeal to a Theorem of Isenberg to show that in the conformal class of the state $(g_{\sharp},\hat{K}_{TT})$ the Lichnerowicz equation can be solved and therefore a CMC state found. Finally we use standard elliptic estimates to show that if the error $\epsilon$ is small enough the CMC state constructed with the conformal method is as close to the states $(g_{H_{i}},-g_{H_{i}})$ and $(g_{G},K_{G})$ (in their respective domains) as we like. 

\vspace{0.2cm}  
{\center \subsubsection{The geometry on a torus neck (the polarized case)}\label{TNPOL}}

On $\field{R}\times \field{R}\times T^{2}$ we look for a (polarized) $T^{2}$-symmetric space-time metric in the coordinates where it looks like
\ben 
g=e^{2a}(-dt^{2}+dx^{2})+Re^{2W}d\theta_{1}^{2}+Re^{-2W}d\theta_{2}^{2}.
\een

\noindent The functions $a,\ R,\ W$ depend on $(t,x)$. Define the coordinates
$(-,+)=(t-x,t+x)$. Derivatives with respect to $-$ and $+$ will be denoted with a subscript $+$ or $-$. In this representation the Einstein equations are equivalent to the system of scalar equations
\begin{equation}\label{eq:R}
\frac{\partial^{2} R}{\partial x^{2}}-\frac{\partial^{2} R}{\partial t^{2}}=0,
\end{equation}
\begin{equation}\label{eq:W}
\frac{\partial}{\partial t}(R\frac{\partial}{\partial t} W)-\frac{\partial}{\partial x}(R\frac{\partial}{\partial x} W)=0,
\end{equation}
\begin{equation}\label{eq:a}
2\frac{R_{\pm}}{R}a_{\pm}=\frac{R_{\pm\pm}}{R}-\frac{1}{2}(\frac{R_{\pm}}{R})^{2}+2W_{\pm}^{2}.
\end{equation}

\noindent Note that equation (\ref{eq:R}) is decoupled from the rest. We make the choice 

\ben
R(x,t)=R_{0}(e^{2(t+x)}+e^{2(t-x)}).
\een

\n The equation (\ref{eq:W}) is the Euler-Lagrange equation of the Lagrangian
\ben
L(t,\partial_{t}W,\partial_{x}W)=\int R(\partial_{t}W)^{2}-R(\partial_{x}W)^{2}dx.
\een

\noindent We make the choice $W(x,t)=W_{1}+W_{0}\arctan e^{2x}$. These solutions are the 
$W$-stable solutions, i.e. those $W$ that with fixed values at the 
boundary (infinity in this case) minimize the potential 
$V=\int R(x,0)(\partial_{x}W)^{2}dx$. We proceed now to find out $a$. Observing that
\ben
2(W_{\pm})^{2}=\frac{W_{0}^{2}}{2\cosh^{2} 2x},
\een

\n equations (\ref{eq:a}) can be written
\begin{equation}\label{eq:a1}
2\frac{R_{\pm}}{R}a_{\pm}=\frac{R_{\pm\pm}}{R}-\frac{1}{2}(\frac{R_{\pm}}{R})^{2}+\frac{W_{0}^{2}}{2\cosh^{2} 2x}.
\end{equation}

\noindent Dividing by $R_{\pm}/R$ and adding and subtracting both equations we get
\ben
\partial_{x}a=-(\frac{1}{2}+\frac{W_{0}^{2}}{2})\tanh 2x,
\een
\ben
\partial_{t}a=\frac{3}{2}+\frac{W_{0}^{2}}{2},
\een

\n which after integration give

\ben
a(x,t)=a(0)-(\frac{1}{2}+\frac{W_{0}^{2}}{2})\frac{1}{2}\ln \cosh 2x  +(\frac{3}{2}+\frac{W_{0}^{2}}{2})t.
\een

\n In the next section we analyze these solutions along some particular space-like foliations.

\vspace{0.2cm}
{\center \subsubsection{The evolution of states on a torus neck}\label{SETN}}

Convene that by {\it observes} we mean a space-like slice ${\mathcal{S}}(t')$ moving with a parametric time $t'$. Let us analyze the solutions found in the previous section with this perspective. First, for those observers that in a forced manner move keeping their $x$-coordinate constant and moving uniformly forward in time $t=t'$, the normalized three-geometry (normalized by $e^{(\frac{3}{2}+\frac{W_{0}^{2}}{2})t}$), collapses along the two-tori into the one-dimensional geometry 
\ben
g_{\infty}=e^{a(0)-(\frac{1}{2}+\frac{W_{0}^{2}}{2})\frac{\ln \cosh 2x}{2}}dx^{2},
\een 

\noindent on the real line and of finite length. However for those observers who freely fall in space-time along time-like geodesics, the normalized three-geometry will be seen to evolve into a hyperbolic cusp
\ben
g_{\infty}=dx^{2}+R_{0}e^{2W_{\pm\infty}}e^{2x}d\theta_{1}^{2}+R_{0}e^{-2W_{\pm\infty}}e^{2x}d\theta_{2}^{2}.
\een

\noindent There are in fact two natural sets of free-falling observers, those which move with positive $x$ and those with negative $x$. Both will observe the 
normalized three-geometry become into hyperbolic cusps (exponentially in time). In between of them the geometry is collapsing, as will be made precise in what follows.

{\it Free falling observers.} We will assume a minor approximation that in no way changes the global picture, nor the precise statements that follow 
on the evolution of the exact geometry. Concentrate on the region $x\geq 10$. On it the metric $g$ (in the $(t,x)$ plane) is almost like
\ben
e^{2((\frac{3}{2}+\frac{W_{0}^{2}}{2})t-(\frac{1}{2}+\frac{W_{0}^{2}}{2})x)}(-dt^{2}+dx^{2}).\een

\noindent We will consider time-like geodesics in this region (towards the increasing direction of $t$). Denote by $s$ their proper time. Then it can be calculated that, independently of the initial velocity, the coordinates $(t(s),x(s))$ of time-like geodesics behave according to
\ben
-(\frac{1}{2}+\frac{W_{0}^{2}}{2})t+(\frac{3}{2}+\frac{W_{0}^{2}}{2})x=\frac{1}{2}\ln \frac{3+W_{0}^{2}}{1+W_{0}^{2}}+
o(\frac{1}{s}),
\een
\ben
-(\frac{1}{2}+\frac{W_{0}^{2}}{2})x+(\frac{3}{2}+\frac{W_{0}^{2}}{2})t=\ln s+\frac{1}{2}\ln \frac{(3+W_{0}^{2})(1+
W_{0}^{2})}{2}+o(\frac{1}{s}).
\een

\noindent What these formulas tells us is that the set of coordinates
\ben
t'=-(\frac{1}{2}+\frac{W_{0}^{2}}{2})x+(\frac{3}{2}+\frac{W_{0}^{2}}{2})t,
\een
\ben
x'=-(\frac{1}{2}+\frac{W_{0}^{2}}{2})t+(\frac{3}{2}+\frac{W_{0}^{2}}{2})x,
\een

\n form the natural coordinate system prescribed by a free-falling set of observers. 
In these new coordinates and after choosing $a(0)=\frac{1}{2}\ln  (-(\frac{1}{2}+\frac{W_{0}^{2}}{2})^{2}+(\frac{3}{2}+\frac{W_{0}^{2}}{2})^{2})$ we get
\ben
g=e^{2t'}(-dt'^{2}+dx'^{2})+R_{0}e^{2(\frac{\pi}{2}W_{0}+
W_{1})}(e^{2(t'+x')}+e^{\frac{2}{2+W_{0}^{2}}(t'-x')})
d\theta_{1}^{2}+\ldots
\een
\ben
\ldots+ R_{0} e^{-2(\frac{\pi}{2}W_{0}+W_{1})}(e^{2(t'+x')}+e^{\frac{2}{2+W_{0}^{2}}
(t'-x')})d\theta_{2}^{2}.
\een

\noindent After making $W_{+\infty}=\frac{\pi}{2}W_{0}+W_{1}$ and normalizing by $e^{2t'}$ 
we see that the local three-geometry exponentially falls into the 
hyperbolic cusp
\ben
g=dx^{2}+R_{0}e^{2W_{+\infty}}e^{2x}d\theta_{1}^{2}+R_{0}
e^{-2W_{+\infty}}e^{2x}d\theta_{2}^{2}
\een

{\center \subsubsection{A convergence-collapse picture}\label{CCP}}

Let us describe now a global foliation of Cauchy hypersurfaces (labeled with a parameter $s\geq 1$) where we can see the picture of convergence-collapse. 
For any $s$ the hypersurface will be defined as: (Zone I) $\{(t,x),\ -(\frac{1}{2}+
\frac{W_{0}^{2}}{2})\ln s+(\frac{3}{2}+\frac{W_{0}^{2}}{2})t=s,\ \mid x\mid \leq 
\ln t\}$, (Zone II)
$\{(t,x),\ s=t'=-(\frac{1}{2}+\frac{W_{0}^{2}}{2})x+(\frac{3}{2}
+\frac{W_{0}^{2}}{2})t,\ x\geq \ln s\}$ and  (Zone III) $\{(t,x),
\ s=t''=(\frac{1}{2}+\frac{W_{0}^{2}}{2})x+(\frac{3}{2}+\frac{W_{0}^{2}}{2})
t,\ x\leq -\ln s\}$. Normalize the three-metrics over the slices 
with the factor $e^{-2s}$. As $s\rightarrow +\infty$ the limit of the normalized three-metrics are: (Zone I)
\ben
g_{\infty}=d\tilde{x}^{2},
\een

\n which is the infinite-length one dimensional geometry on the real line, and (Zone II)
\ben
g_{\infty}=dx^{2}+R_{0}e^{2W_{+\infty}}e^{2x}d\theta_{1}^{2}+R_{0}e^{-2W_{+\infty}}e^{2x}d\theta_{2}^{2},
\een

\begin{figure}[h]
\centering
\includegraphics[width=11cm,height=7.5cm]{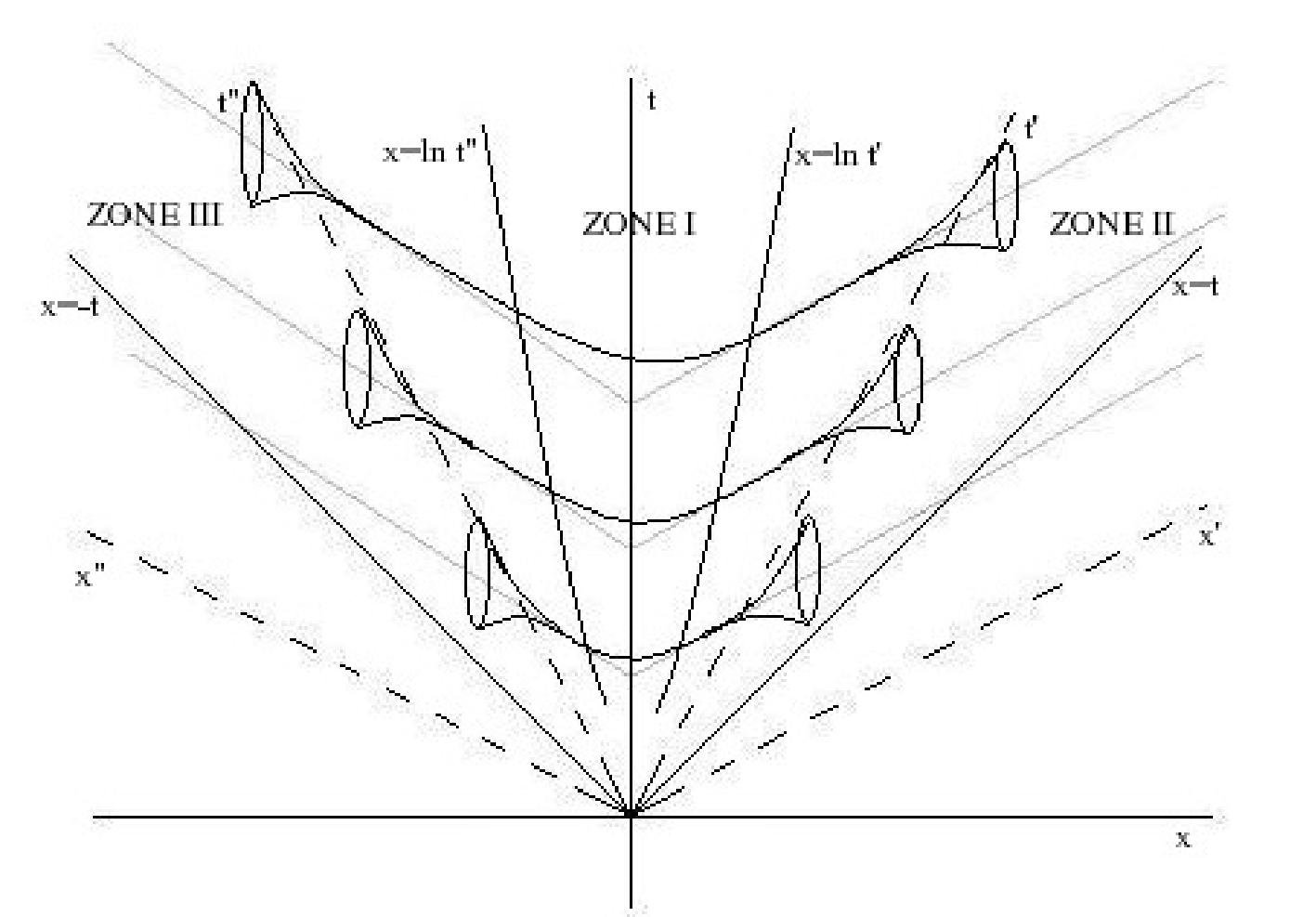}
\caption[Convergence-collapse in a torus neck]{A schematic figure showing 
the evolution of the normalized three-geometry.}\label{fig:Fig4}
\end{figure}

\n on the whole $\field{R}\times T^{2}$, and similarly for the Zone III. A schematic picture can be seen in Figure \ref{fig:Fig4}. 

\vspace{0.5cm}

{\center \subsubsection{The geometry on a torus neck (the non-polarized case)}\label{GTNNP}}

In this section we follow the same strategy as in Section \ref{TNPOL} to find (Gowdy) $T^{2}$-symmetric space-time solutions but this time non-polarized. On 
$\field{R}\times \field{R}\times T^{2}$ we look 
for a non-polarized $T^{2}$-symmetric metric in the coordinates where it looks like
\beq\label{eeeq1}
g=e^{2a}(-dt^{2}+dx^{2})+R(e^{2W}+q^{2}e^{-2W})d\theta_{1}^{2}-Rqe^{-2W}2d\theta_{1}d\theta_{2}+
Re^{-2W}d\theta_{2}^{2},
\eeq

\noindent and where $a,\ R,\ W$ depend only on $(t,x)$ or $(u,v)=(-,+)=(t-x,t+x)$. In this representation the Einstein equations reduce to
\beq
R_{+-}=0,
\eeq
\begin{equation}\label{eeeq2}
2\frac{R_{++}}{R}-(\frac{R_{+}}{R})^{2}+4W_{+}^{2}+q_{+}^{2}e^{-4W}-4a_{+}\frac{R_{+}}{R}=0,
\end{equation}
\begin{equation}\label{eeeq3}
2\frac{R_{--}}{R}-(\frac{R_{-}}{R})^{2}+4W_{-}^{2}+q_{-}^{2}e^{-4W}-4a_{-}\frac{R_{-}}{R}=0,
\end{equation}
\begin{equation}\label{eeeq4}
(RW_{-})_{+}+(RW_{+})_{-}+Rq_{+}q_{-}e^{-4W}=0,
\end{equation}
\begin{equation}\label{eeq5}
(Re^{-4W}q_{+})_{-}+(Re^{-4W}q_{-})_{+}=0.
\end{equation}

\noindent Again we make the choice $R(x,t)=R_{0}e^{2t}\cosh(2x)$. With this choice 
we will solve for time-independent $W$ and $q$ realizing arbitrary flat 
metrics on the two tori at the ends, i.e. which have prescribed asymptotic $q_{\infty},\ q_{-\infty},\ W_{\infty},\ W_{-\infty}$. After that we will solve for $a$. 

{\it Solving for time independent $W$ and $q$.} Equation (\ref{eeq5}) forces $q'$ to satisfy
\begin{equation}\label{eeeq6}
q'=\frac{2ce^{4W}}{\cosh(2x)}.
\end{equation}

\n where $c$ is an arbitrary constant. With $q'$ of this form, equation (\ref{eeeq4}) forces $W$ to satisfy
\begin{equation}\label{eeeq7}
W''+2\tanh(2x)W'=\frac{-2c^{2}e^{4W}}{\cosh^{2}(2x)},
\end{equation}

\n The strategy to find the solutions to (\ref{eeeq6})-(\ref{eeeq7}) for $W$ and $q$ and having prescribed asymptotic values at the ends (i.e. when $x\rightarrow \pm \infty$) is the following. Fix $c$ first. Then find $W$ having the prescribed asymptotic values $W(\infty)=W_{\infty}$ and $W(-\infty)=W_{-\infty}$. Then vary 
$c$ keeping fixed the asymptotic conditions for $W$ and prove that we can reach at some $c$ the prescribed asymptotic value $q(\infty)=q_{\infty}$ if $q(-\infty)=q_{-\infty}$ was prescribed. We will accomplish that by proving that varying $c$ from some value $c_{0}$ toward zero, the integral from $-\infty$ to $\infty$ of equation (\ref{eeeq6}) that defines $q(\infty)$ reaches (having $q_{-\infty}$ as the lower limit of integration prescribed) all possible values. Although equation (\ref{eeeq7}) is highly non-linear, it can be integrated exactly. We note that equation (\ref{eeeq7}) is equivalent (unless $W$ is constant in which case $c=0$ and $q$ is constant) to
\begin{equation}\label{eq8}
((\cosh (2x)W')^{2})'=-(c^{2}e^{4W})',
\end{equation}

\n which gives
\begin{equation}\label{eq9}
\cosh^{2} (2x)W'^{2}=-c^{2}e^{4W}+A^{2},
\end{equation}

\n for $A>0$, an arbitrary positive constant. Taking the square root of (\ref{eq9}) we get a separable variables ODE. After integration we get
\begin{equation}\label{eeq10}
W=-\frac{1}{2}\ln \frac{\mid c\mid}{A}\cosh (-2A\arctan e^{2x}+B),
\end{equation}

\n with $B$ and arbitrary constant. We need to find $A$ and $B$ that solve the asymptotic conditions for $W$ i.e.
\ben
\frac{\mid c\mid}{A}\cosh B=e^{-2W_{-\infty}},
\een
\ben
\frac{\mid c\mid}{A}\cosh(-\pi A+B)=e^{-2W_{\infty}}.
\een

\noindent Making the change of variables $A=\frac{B-D}{\pi}$ we get the equivalent equations
\begin{equation}\label{eeq13}
B=D+\pi \mid c\mid e^{2W_{\infty}}\cosh D,
\end{equation}
\begin{equation}\label{eeq14}
D=B-\pi \mid c\mid e^{2W_{-\infty}}\cosh B.
\end{equation}

\begin{figure}[h]
\centering
\includegraphics[width=80mm,height=90mm]{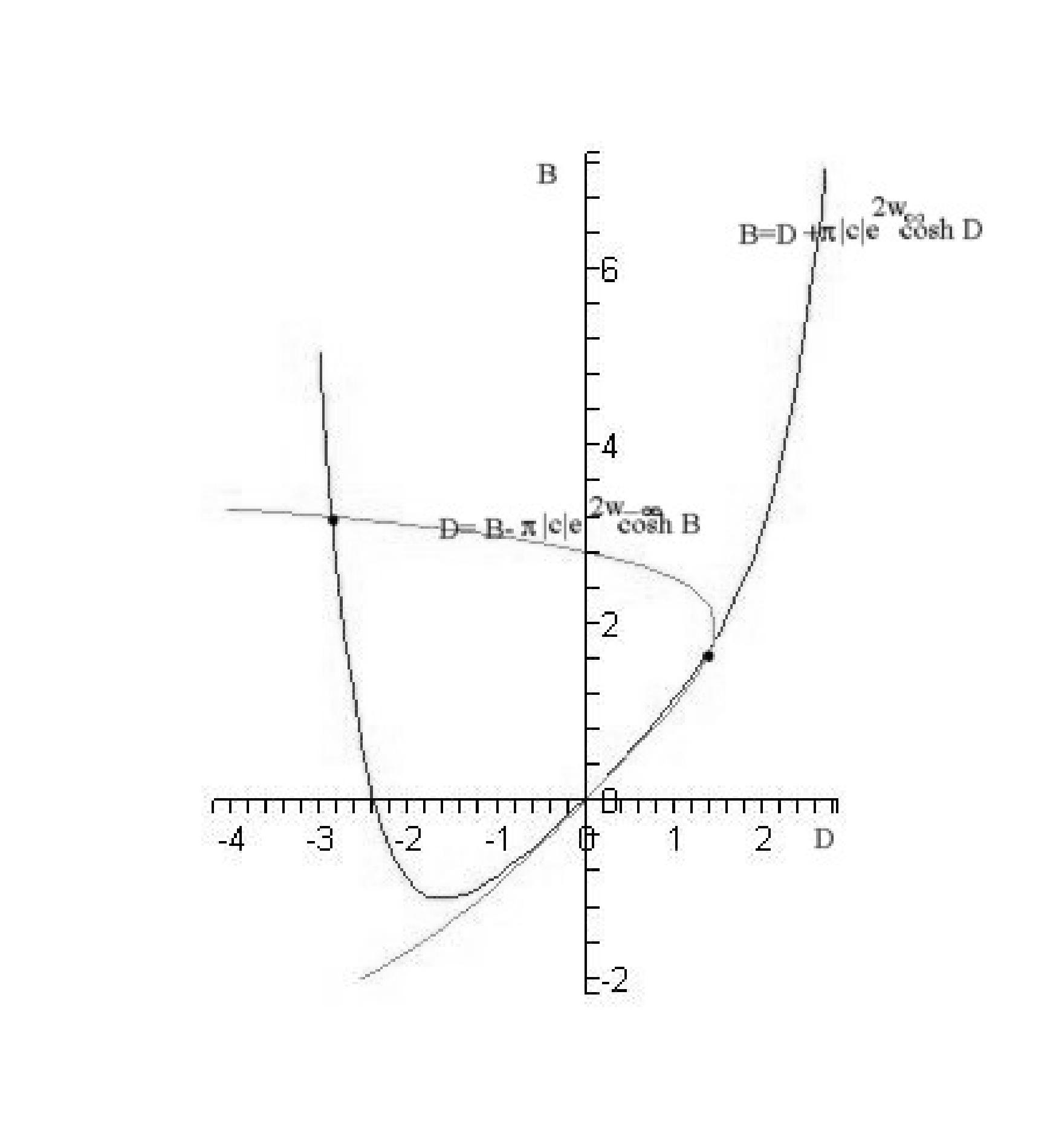}
\caption{The graphs of $B(D)$ (from (\ref{eeq13})) and $D(B)$ (from (\ref{eeq14})) for a 
small $c$.}\label{fig:BD}
\end{figure}

\noindent Now the problem is to understand the solutions $B$ and $D$ to (\ref{eeq13})-(\ref{eeq14}) as functions of $c$, $W_{\infty}$ and $W_{-\infty}$. If we graph $B(D)$ (from (\ref{eeq13})) and $D(B)$ (from (\ref{eeq14})) on the same $B-D$-coordinates axis, 
we see (observe the factor $|c|$ in front of $\cosh D$ and $\cosh B$) that there is some positive $c_{0}$ above which there are no 
solutions (the graphs do not intersect), at which there is only one and below which there are only two solutions. See Figure \ref{fig:BD}.  

In the following we will analyze the solutions $A$ and $B$ as $c\rightarrow 0$. We will see that given a prescribed value $q_{-\infty}$ 
we get any asymptotic value for $q_{\infty}$ by varying $c$ from $c_{0}$ towards zero.
The equation
\ben
e^{2W_{-\infty}}\cosh B=e^{2W_{\infty}}\cosh D,
\een

\n gives for the each one of the two different branches (of solutions ($B$, $D$)) the following behaviors
\begin{enumerate}
\item (Branch I). Either $W_{\infty}=W_{-\infty}$ for which we get (observe that $A=B-D>0$)
\ben
\begin{array}{l}
B=-D\rightarrow  0,\\
\frac{\mid c\mid}{A}\rightarrow e^{-2W_{-\infty}},
\end{array}
\een

\n or $W_{\infty}\neq W_{-\infty}$ for which we get
\ben
\begin{array}{l}
B\rightarrow \infty \ if\  
W_{\infty}>W_{-\infty}\ (or\ -\infty \ if\ W_{\infty}<W_{-\infty}),\\
B-D\rightarrow 2(W_{\infty}-W_{-\infty})\ (or\ -2(W_{\infty}-W_{-\infty})),\\
A\rightarrow \frac{2}{\pi}(W_{\infty}-W_{-\infty})\ (or\ -\frac{2}{\pi}(W_{\infty}-W_{-\infty})).
\end{array}
\een

\item (Branch II) For any $W_{\infty}, W_{-\infty}$
\ben
\begin{array}{l}
B\rightarrow \infty,D\rightarrow -\infty,\\
B+D\rightarrow 2(W_{\infty}-W_{-\infty}),\\
A\sim \frac{2B-2(W_{\infty}-W_{-\infty})}{\pi}.
\end{array}
\een
\end{enumerate}

\noindent With these behaviors for $A$ and $B$ (as $c\rightarrow 0$) we get

(Branch I). The formula for $q'$
\ben
q'=\frac{c}{(\cosh(2x))(\frac{\mid c\mid}{A}\cosh(-2A\arctan e^{2x}+B))^{2}},
\een

\n shows that, starting at an arbitrary $q_{-\infty}$, the function $q$ approaches (uniformly) to the constant function $q=q_{-\infty}$.

(Branch II). The formula for $q'$ approximates to
\ben
q'\sim\frac{\pm e^{-2W_{-\infty}}(2B-2(W_{\infty}-W_{-\infty}))}{\pi\cosh B\cosh (2x) (e^{-2W_{-\infty}}(\cosh B)^{-1}\cosh (-2A\arctan 
e^{2x}+B))^{2}}.
\een

\n Rearranged it reads
\begin{equation}\label{eeq20}
q'\sim\frac{\pm e^{2W_{-\infty}}(2B-2(W_{\infty}-W_{-\infty}))\cosh B}{\pi\cosh (2x)\cosh (-2A\arctan e^{2x}+B))^{2}}.
\end{equation}

\n The factor 
\ben
\cosh(-2A\arctan e^{2x}+B)=\cosh(B(-2\frac{A}{B}\arctan e^{2x}+1)),
\een

\n in the denominator of equation (\ref{eeq20}), can be bounded above in the interval 
$-1\leq x\leq 1$ by
\ben
\cosh 2Bx.
\een

\n (To see that note that $-2\frac{A}{B}\rightarrow \frac{-4}{\pi}$, linearize $\arctan e^{2x}$ ($x\sim 0$) and get the bound). The integral
\ben
\pm \int_{-1}^{1}\frac{e^{-2W_{-\infty}}(2B-2(W_{\infty}-W_{-\infty}))\cosh B}{\cosh 2x (\cosh 2Bx)^{2}}dx,
\een

\n is equal, after the change of variables $Bx=u$, to
\ben
\pm \int_{-B}^{B}\frac{e^{2W_{-\infty}}(2B-2(W_{\infty}-W_{-\infty}))\cosh B}{B\cosh \frac{2u}{B}\cosh^{2}2u}du,
\een

\n that clearly diverges to $\pm$ infinity as $B$ goes to infinity.

{\it Solving for $a$.} To find out the expression for $a$ we follow the same procedure as in the polarized case. 
We find $\dot{a}$ and $a'$ from equations 
(\ref{eeeq2}) and (\ref{eeeq3}) and then integrate in time $(t)$ and space $(x)$. As 
$W$ and $q$ are time independent we have
\ben
4W_{\pm}^{2}+q_{\pm}^{2}e^{-W}=W'^{2}+\frac{q'^{2}}{4}e^{-4W}.
\een

\noindent Equation (\ref{eq9}) gives 
\ben
W'^{2}+\frac{q'^{2}}{4}e^{-4W}=\frac{A^{2}}{\cosh^{2}2x}.
\een

\n This formula makes equations (\ref{eeeq2}) and (\ref{eeeq3}) to have the same form as equations 
(\ref{eq:a1}) but with $W_{0}^{2}$ replaced by 
$\frac{A^{2}}{2}$. This gives the following expression for $a$
\ben
a(x,t)=a(0)-(\frac{1}{2}+\frac{A^{2}}{4})\frac{1}{2}\ln \cosh 2x  +(\frac{3}{2}+\frac{A^{2}}{4})t.
\een

\noindent The analysis of the convergence-collapse picture for these non-polarized solutions follows exactly as in the polarized case.

{\center \subsubsection{The gluing}\label{GL}}

{\it CMC states in a torus neck.} For simplicity we will work with the polarized solution in the torus neck we have found before The computations carry over to the non-polarized case as well. We will find a CMC
slice, $t=s(x)$, of the solution
\ben
g=e^{2a}(-dt^{2}+dx^{2})+Re^{2W}d\theta_{1}^{2}+Re^{-2W}d\theta_{2}^{2},
\een
\ben
\ \ \ a(x,t)=a(0)-(\frac{1}{2}+\frac{W_{0}^{2}}{2})\frac{1}{2}\ln \cosh 2x  +
(\frac{3}{2}+\frac{W_{0}^{2}}{2})t,
\een
\ben
\ \ \ R(x,t)=R_{0}(e^{2(t+x)}+e^{2(t-x)}),
\een
\ben
\ \ \ W(x,t)=W_{1}+W_{0}\arctan e^{2x},
\een

\n with $k=-3$ and asymptotically of the form $t=s(x)\sim t_{0}\pm\frac{(1+W_{0}^{2})}{(3+W_{0}^{2})}x$. With this asymptotic we guarantee having (almost) flat cone initial states on the ends. The way to find such CMC slice is by
finding appropriate barriers. To do that we first find a general expression for
the mean curvature of a general section $t=s(x)$. We keep the discussion brief. Given a slice
$t=s(x)$ introduce a coordinate system $(\bar{x},\bar{t},\bar{\theta}_{1},\bar{\theta}_{2})$
defined as
\ben
x=\bar{x}+s'(\bar{x})\bar{t},
\een
\ben
t=s(\bar{x})+\bar{t},
\een
\ben
\theta_{1}=\bar{\theta_{1}},
\een
\ben
\theta_{2}=\bar{\theta_{2}}.
\een

\noindent In these coordinates the metric $g$ is written 
\ben
g=-\bar{N}^{2}d\bar{t}^{2}+\bar{g}(d\bar{x}+\bar{X}d\bar{t})(d\bar{x}+\bar{X}d\bar{t})+
Re^{2W}d\bar{\theta}_{1}^{2}+Re^{-2W}d\bar{\theta}_{2}^{2},
\een

\n where
\ben
\bar{g}=e^{2a}((1+s''\bar{t})^{2}-s'^{2}),
\een
\ben
\bar{N}^{2}=e^{2a}(1-s'^{2}).
\een

\noindent and $\bar{X}=0$ when $\bar{t}=0$. From this $k$ is calculated (at the slice 
$t=s(x)$) as
\ben
k=-\frac{1}{e^{a}\sqrt{1-s'^{2}}}(\partial_{\bar{t}}a+\frac{s''}{1-s'^{2}}+\frac{\partial_{\bar{t}}R}{R}),
\een

\n where
\ben
\partial_{\bar{t}}a=\partial_{t}a+s'\partial_{x}a=-(\frac{1}{2}+\frac{W_{0}^{2}}{2})s'\tanh 2x+(\frac{3}{2}+\frac{W_{0}^{2}}{2}),
\een
\ben
\frac{\partial_{\bar{t}}R}{R}=\frac{\partial_{x}Rs'+\partial_{t}R}{R}=2s'\tanh 2x +2,
\een

\n which gives
\ben
k(x)=-\frac{1}{\sqrt{1-s'^{2}}}e^{f}
(\frac{s''}{1-s'^{2}}-(\frac{1}{2}+\frac{W_{0}^{2}}{2})s'\tanh 2x+(\frac{3}{2}+\frac{W_{0}^{2}}{2})+2s'\tanh 2x +2),
\een

\n with
\ben
f=-(a(0)-(\frac{1}{2}+\frac{W_{0}^{2}}{2})\frac{1}{2}\ln\cosh 2x+(\frac{3}{2}+\frac{W_{0}^{2}}{2})s).
\een

\begin{Remark}{\rm Note that $k(s(x)+\tau)=e^{-(\frac{3}{2}+\frac{W_{0}^{2}}{2})\tau}k(s(x))$. This implies in particular that once we have obtained a CMC slice a CMC foliation is obtained by shifting it in the ($t$) time direction.}
\end{Remark}

\noindent Now, to construct the barriers, note that for the section $t=s(x)=t_{0}+(\frac{1+W_{0}^{2}}{3+W_{0}^{2}})x$, $k$ is asymptotically (i.e. as $x\rightarrow +\infty$) constant. A direct calculation shows that for the pair of sections (on the right end)
\begin{equation}\label{s+-}
t=s_{\pm}(x)=t_{0}+\frac{1+W_{0}^{2}}{3+W_{0}^{2}}x\pm\frac{1}{x},
\end{equation}

\n the asymptotic (to leading terms) is
\ben
-k\sim-k_{0}e^{\mp(\frac{3}{2}+\frac{W_{0}^{2}}{2})\frac{1}{x}}(1+O(\frac{1}{x})).
\een

\n The last formula shows that $-k(s_{+})<-k_{0}<-k(s_{-})$ asymptotically. The extension of those sections
to the center of the neck can be carried as follows. Take two sections symmetric with respect to the $t$-axis, that (say on the 
right) are i. any smooth section ($s_{+}$) from $0$ to $10$ with $s''>0$ and $s_{+}(10)+\frac{1+W_{0}^{2}}{3+W_{0}^{2}}(x-10)-\ln 
(x-9)$ thereafter ii. any smooth section ($s_{-}$) from $0$ to $10$ with $s''>0$ and equal to $s_{-}(10)+\frac{1+W_{0}^{2}}{3+W_{0}^{2}}(x-10)+
\ln (x-9)$ thereafter. It is easy to see using the Remark above that by shifting the section $s_{-}$ upwards, at some shift the sections 
have disjoint range of their mean curvatures (between the points of intersection) and that at the point of intersection their tangents 
are $\frac{1+W_{0}^{2}}{3+W_{0}^{2}}$ up to $\sim 1/x$. Due to that, it is easy to continue these two sections as was said above (in equation (\ref{s+-})), starting 
from an $x$ slightly less than the $x$ where they intersect, in such a way that they have disjoint range of mean curvatures but asymptotically 
approaching to  $s(x)=t_{0}+\frac{1+W_{0}^{2}}{3+W_{0}^{2}}$.

Note that given a CMC slice as was described above, the same slice is CMC with the same mean curvature if on the metric $g$ we replace
$R_{0}$ by $R_{0}e^{-2\delta}$. Also note that on the $(x',t')$ coordinates, for large $x'$
the metric is written approximately
\ben
\begin{split}
g=&e^{2t'}(-dt'^{2}+dx^{2})+R_{0}e^{2(\frac{\pi}{2}W_{0}+
W_{1})}(e^{2(t'+x')}+e^{\frac{2}{2+W_{0}^{2}}(t'-x')})
d\theta_{1}^{2}\\
&+ R_{0}
e^{-2(\frac{\pi}{2}W_{0}+W_{1})}(e^{2(t'+x')}+e^{\frac{2}{2+W_{0}^{2}}
(t'-x')})d\theta_{2}^{2}.
\end{split}
\een

\n Thus, changing $R_{0}$ by $R_{0}e^{-2\delta}$ and changing the $x'$ coordinate by $x''=x'-\delta$
the metric approximates to any given desired order to the flat cone state,
\ben
\begin{split}
g=&e^{2t'}(-dt'^{2}+dx''^{2})+R_{0}e^{2(\frac{\pi}{2}W_{0}+
W_{1})}e^{2(t'+x'')}d\theta_{1}^{2}\\
&+ R_{0}
e^{-2(\frac{\pi}{2}W_{0}+W_{1})}e^{2(t'+x'')}d\theta_{2}^{2}.
\end{split}
\een

\n Note moreover that the distance between standard parts on the cusps get increased by $\sim 2\delta$. $\delta$ therefore parametrizes 
the family of CMC initial states displaying a convergence-collapse picture.

{\it A traceless transverse tensor.} Having now a metric on the torus neck
we glue it to the hyperbolic metrics $dx^{2}+e^{2x}g_{T_{i}}$ (on the right (say $i=2$) and the left (say $i=1$) of the neck) along intervals
of left one around $x=x_{0}$ and preserving the $T^{2}$ symmetry. There is some freedom of course in this process. We will use it in a moment. We will look for a $T^{2}$-symmetric transverse traceless
$(2,0)$-tensor $\hat{K}_{TT}$ with respect to the metric that resulted from the gluing. 
Moreover we will demand the components of $\hat{K}_{TT}$ to be zero except for $\hat{K}_{TT,xx}$, $\hat{K}_{TT,\theta_{1}\theta_{1}}$ and $\hat{K}_{TT,\theta_{2}\theta_{2}}$.
Finally we demand $\hat{K}_{TT}$ to be unchanged on the region inside the neck which is not the gluing region and, similarly, we demand $\hat{K}_{TT}$ to be unchanged inside the bulk of the hyperbolic
manifolds $H_{1}$ and $H_{2}$ which is not the gluing region. Thus we want $\hat{K}_{TT}$
to be zero on the hyperbolic sector and right after the gluing. Observing that for any 
$T^{2}$-symmetric metric the connection coefficients 
$\Gamma_{\theta_{i}\theta_{j}}^{\theta_{k}}$ for $i,j,k$ equal to $0$ or $1$ are zero and similarly for $\Gamma_{x\theta_{i}}^{x}$ and $\Gamma^{\theta_{i}}_{xx}$ for $i=0,1$ we have
\ben
\nabla_{i}\hat{K}_{TT,\theta_{j}}^{i}=0,\ j=0,1.
\een

\n For $\nabla_{i}\hat{K}_{TT\ x}^{i}$ we compute
\beq\label{eq:d}
\nabla_{i}\hat{K}_{TT,x}^{i}=\partial_{x}\hat{K}_{TT,x}^{x}+(\Gamma_{x\theta_{2}}^{\theta_{2}}-\Gamma_{x\theta_{1}}^{\theta_{1}})\hat{K}_{TT,x}^{x}
\eeq

\n where we have implicitly used that $\hat{K}_{TT,x}^{x}+
\hat{K}_{TT,\theta_{1}}^{\theta_{1}}+\hat{K}_{TT,\theta_{2}}^{\theta_{2}}=0$. We need to find a solution of (\ref{eq:d}) being exactly zero after an interval of length one. To do that we
choose the glued metric in such a way that $\Gamma^{\theta_{1}}_{x\theta_{1}}\neq\Gamma^{\theta_{2}}_{x\theta_{2}}$
(with a small difference) on an interval of length one half inside the gluing interval. Then choose $\hat{K}^{\theta_{1}\theta_{1}}$
such that the solution to (\ref{eq:d}) is exactly zero right after the gluing region. One can check that this can be done using the integral formula for the solution of a first order ODE.\\

{\it Estimates.} Once having $(g,K)$ with $div K=0$ and $tr_{g}K=k$ we invoke a theorem of Isenberg \cite{I} guaranteeing that the Lichnerowicz equation is solvable as long as $\hat{K}\neq 0$ and $k\neq 0$ as is our case. To estimate the solution to the Lichnerowicz equation
\ben
\Delta \phi=\frac{1}{8}R_{g}\phi-\frac{1}{8}|\hat{K}|^{2}_{g}\phi^{-7}+\frac{k^{2}}{12}\phi^{5},
\een

\n we use the maximum principle and the standard local elliptic estimates. From the maximum principle we get
\ben
R_{g}\phi(x_{max})-|\hat{K}|^{2}\phi(x_{max})^{-7}+\frac{k^{2}}{12}\phi(x_{max})^{5}\leq 0,
\een

\n Now note that $R_{g}=|\hat{K}|^{2}-\frac{2}{3}k^{2}+\epsilon(x)$ where $\epsilon(x)$ is
nonzero only on the gluing region. Using this in the last equation gives
\beq\label{eps}
|\hat{K}|^{2}(\phi(x_{max})-\phi^{-7}(x_{max}))+\frac{2}{3}k^{2}(\phi(x_{max})^{5}-\phi(x_{max}))+\epsilon(x_{max})\phi(x_{max})\leq 0.
\eeq

\n Observe that $\|K\|_{L^{\infty}_{g}}$ is bounded with a bound independent of $\epsilon$. We see from equation (\ref{eps}) that when $\|\epsilon\|_{L^{\infty}}\rightarrow 0$ 
then $\|\phi-1\|_{L^{\infty}}\rightarrow 0$. Standard elliptic estimates show that in fact 
$\|\phi-1\|_{C^{2,\alpha}}\rightarrow 0$.

{\center \section{Long time geometrization of the Einstein flow}\label{LTGEF}}

{\center \subsection{The long-time geometrization of the Einstein flow}\label{LTGEF2}}

In this section we prove the following Theorem.

\vspace{0.2cm}
\begin{T}\label{LTGFT} Let $\Sigma$ be a compact three-manifold with $Y(\Sigma)\leq 0$. Say $(\tilde{g},\tilde{K})(\sigma)$ is a cosmologically normalized flow with $\tilde{\Ef}(\sigma)\leq \La$
where $\La$ is a positive constant. Then, the cosmologically normalized flow $(\tilde{g},\tilde{K})(\sigma)$ persistently geometrizes the manifold $\Sigma$. Moreover the induced geometrization is the Thurston geometrization iff ${\Nu}(\sigma)\downarrow \Nu_{inf}=(-\frac{1}{6}Y(\Sigma))^{\frac{3}{2}}$.
\end{T}

We need some preliminary propositions.

\begin{Prop}\label{ELEM} Let $\Sigma$ be a compact three-manifold. Say $g_{0}$ is a $H^{2}$-Riemannian-metric on $\Sigma$. Say $p\in \Sigma$ and $2R<r_{2}(p)$ where $r_{2}(p)$ is the $H^{2}$-harmonic radius of the metric $g_{0}$ at the point $p$. According to the definition of $H^{2}$-harmonic radius we consider a harmonic coordinate system $\{x\}$ covering $B_{g_{0}}(p,r_{2}(p))$ and satisfying 
\begin{equation}\label{cc}
\frac{3}{4}\delta_{jk}\leq g_{0,jk}\leq \frac{4}{3}\delta_{jk},
\end{equation}
\begin{equation}\label{ccc}
r_{2}(p)(\sum_{|I|=2,j,k}\int_{B_{g_{0}}(p,r_{2}(p))}|\frac{\partial^{I}}{\partial x^{I}}g_{jk}|^{2}dv_{x})\leq 1.
\end{equation}

\n Then there is $\epsilon(R)$ such that if $\|g-g_{0}\|_{H^{2}_{\{x\}}(B_{g_{0}}(p,R))}\leq \bar{\epsilon}\leq \epsilon(R)$ the inclusions $id:H^{i}_{g}(B_{g_{0}}(p,R))\hookrightarrow H^{i}_{g_{0}}(B_{g_{0}}(p,R))$ and $id:H^{i}_{g_{0}}(B_{g_{0}}(p,R))\hookrightarrow H^{i}_{g}(B_{g_{0}}(p,R))$ for $i=0,1,2$ have norms controlled by $\bar{\epsilon}$ and $R$.

\end{Prop}

\n {\bf Proof:} 

Note first the Sobolev embeddings\footnote{It is crucial that the embeddings are from $H^{\star}_{\{x\}}(B_{g_{0}}(p,R))$ and not from $H^{\star}_{0,\{x\}}(B_{g_{0}}(p,R))$.
This is justified by the fact that, in the coordinate system $\{x\}$ the set $B_{g_{0}}(p,R)$ has the cone property at its boundary (see \cite{GT}, pg. 158).}
\beq\label{S1}
H^{1}_{\{x\}}(B_{g_{0}}(p,R))\hookrightarrow L^{4}_{\{x\}}(B_{g_{0}}(p,R)),
\eeq
\beq\label{S2}
H^{2}_{\{x\}}(B_{g_{0}}(p,R))\hookrightarrow C^{0}_{\{x\}}(B_{g_{0}}(p,R)).
\eeq

\n From (\ref{S1}) we see that $\|g-g_{0}\|_{C^{0}_{\{x\}}(B_{g_{0}}(p,R))}\leq \epsilon'(\bar{\epsilon},R)$ with $\epsilon'\rightarrow 0$ as $\bar{\epsilon}\rightarrow 0$ (and $R$ fixed). This in particular implies that 
\ben
C_{1}g_{0,ij}\leq g_{ij}\leq C_{2}g_{0,ij},
\een

\n where $C_{1}$ and $C_{2}$ depend on $\bar{\epsilon}$ and $R$ and tend to one as $\bar{\epsilon}\rightarrow 0$ (keeping $R$ fixed). This proves the inequality
\ben
C_{1}\|U\|_{L^{2}_{g_{0}}(B_{g_{0}}(p,R))}\leq \|U\|_{L^{2}_{g}(B_{g_{0}}(p,R))}\leq C_{2}\|U\|_{L^{2}_{g_{0}}(B_{g_{0}}(p,R)},
\een

\n for some $C_{1}$ and $C_{2}$ dependent on $\bar{\epsilon}$ and $R$, which terminates the case $i=0$. In the following we will use the notation $C_{1}$, $C_{2}$ to denote generic quantities depending on $\bar{\epsilon}$ and $R$. Let us prove the case $i=1$ now. Denote by $\nabla$ and $\bar{\nabla}$ the covariant derivatives associated to $g_{0}$ and $g$ respectively. Write $\bar{\nabla}=\nabla +\Gamma$. With this notation we have
\ben
|\bar{\nabla} U|^{2}_{g}=|\nabla U +\Gamma*U|^{2}_{g}\leq C_{2}(|\nabla U|^{2}_{g_{0}}+|\Gamma|^{2}_{g_{0}}|U|^{2}_{g_{0}}).
\een

\n Integrating we get
\beq\label{yoque}
\begin{split}
\int_{B_{g_{0}}(p,R)}|\bar{\nabla} U|^{2}_{g}dv_{g}\leq C_{2}(&\int_{B_{g_{0}}(p,R)}|\nabla U|^{2}_{g_{0}}dv_{g_{0}}\\
&+(\int_{B_{g_{0}}(p,R)}|\Gamma|^{4}_{\{x\}}dv_{x})^{\frac{1}{2}}(\int_{B_{g_{0}}(p,R))}|U|^{4}_{g_{0}}dv_{g_{0}})^{\frac{1}{2}}).
\end{split}
\eeq

\n It is direct to see from the formula
\ben
\Gamma^{k}_{ij}=\frac{1}{2}(\nabla_{i}(g_{jm}-g_{0,jm})+\nabla_{j}(g_{im}-g_{0,im})-\nabla_{m}(g_{ij}-g_{0,ij}))g^{km},
\een

\n that $\|\Gamma\|_{H^{1}_{\{x\}}(B_{g_{0}}(p,R))}\rightarrow 0$ as $\bar{\epsilon}\rightarrow 0$. Sobolev embeddings applied to equation (\ref{yoque}) give
\ben
\|\bar{\nabla}U\|^{2}_{L^{2}_{g}(B_{g_{0}}(p,R))}\leq C_{2}\|U\|^{2}_{H^{1}_{g_{0}}(B_{g_{0}}(p,R))},
\een

\n and thus
\ben
\|U\|^{2}_{H^{1}_{g}(B_{g_{0}}(p,R))}\leq C_{2}\|U\|^{2}_{H^{1}_{g_{0}}(B_{g_{0}}(p,R))},
\een

\n as desired. Let us prove the other inequality. Write
\ben
|\nabla U|_{g_{0}}^{2}=|\bar{\nabla}U-\Gamma*U|^{2}_{g_{0}}\leq C_{1}(|\bar{\nabla} U|^{2}_{g}+|\Gamma|^{2}_{g_{0}}|U|^{2}_{g_{0}}).
\een

\n Integrating we get
\ben
\int_{B_{g_{0}}(p,R)}|\nabla U|^{2}_{g_{0}}dv_{g_{0}}\leq C_{1}(\int_{B_{g_{0}}(p,R)}|\bar{\nabla} U|^{2}_{g}dv_{g}+(\int_{B_{g_{0}}(p,R)}|\Gamma|^{4}_{g_{0}}dv_{g_{0}})^{\frac{1}{2}}(\int_{B_{g_{0}}(p,R)}|U|^{4}_{g_{0}})^{\frac{1}{2}}),
\een

\n Again Sobolev embeddings give
\ben
\|\nabla U\|_{L^{2}_{g_{0}}(B_{g_{0}}(p,R))}^{2}\leq C_{1}(\|\bar{\nabla}U\|^{2}_{L^{2}_{g}(B_{g_{0}}(p,R))}+\|\Gamma\|_{H^{1}_{\{x\}}(B_{g_{0}}(p,R))}^{2}\|U\|^{2}_{H^{1}_{g_{0}}(B_{g_{0}}(p,R))}).
\een

\n Moving the second term on the right hand side to the left side and choosing $\bar{\epsilon}$ sufficiently small\footnote{Note that $C_{1}$ does not blow up as $\bar{\epsilon}\rightarrow 0$.} we have
\ben
\|U\|_{H^{1}_{g_{0}}(B_{g_{0}}(p,R))}\leq C_{1}\|U\|_{H^{1}_{g}(B_{g_{0}}(p,R))},
\een

\n as desired. The case $i=2$ follows easily from the case $i=1$.\ep

We consider now the Einstein flow with zero shift, i.e. we assume we have set $X=0$.

\begin{Prop}\label{CF}{\rm (Continuity of the flow.)} Say $\Sigma$ is a compact three-manifold with $Y(\Sigma)\leq 0$. Say $(g,K)(k)$ is a long-time Einstein flow with domain (at least) $[-3,0)$. Suppose that $\Ef(k)\leq \Lambda$ where $\La$ is a positive constant. We use the notation $(g_{0},K_{0})=(g(-3),K(-3))$, $k_{0}=-3$ and $\Nu(-3)=\Nu_{0}$. Say $p\in \Sigma$ and $r_{2,g_{0}}(p)\geq 2R$. Then for any $\epsilon>0$ there is $\delta k(\La, \Nu_{0},R)>0$ such that 
\ben
\sup_{k\in [k_{0},k_{0}+\delta k]}\{\|(g,K)(k)-(g,K)(k_{0})\|_{H^{2}_{g_{0}}(B_{g_{0}}(p,R))\times H^{1}_{g_{0}}(B_{g_{0}}(p,R))}\}\leq \epsilon . 
\een

\end{Prop}

\begin{Remark}. {\rm i. Proposition \ref{CF} would be self evident (see \cite{Rei3}) if we have a priori control on $r_{2,g_{0}}$ over the whole manifold $\Sigma$. It is not a priori clear how is that the regions where the harmonic radius (or volume radius) is small may affect the evolution of the regions where it is not, even in the short time. What Proposition \ref{CF} shows is that under an a priori bound in $\Ef$ this influence is not noticeable in a definite interval of time $t=k$ (depending on $\Ef$, $\nu_{0}$ and $R$). Note however that
we do not make any claim about the continuity in $H^{2}_{g_{0}}(B_{g_{0}}(p,R))$ of the lapse $N$. As we will remark later the $H^{2}_{\g_{0}}(B_{g_{0}}(p,R))$ norm of $N$ is indeed controlled but we do not know whether $N$ satisfies a continuity of the type claimed for $g$ and $K$ (in their respective spaces). In particular we do not have any estimation (in any norm) of the time derivative of $N$ on $B_{g_{0}}(p,R)$ even for short times. This issue will appear later in Proposition \ref{KRIEBB}.

ii. The Proposition \ref{CF} is evidently true if we use the cosmologically normalized variables $(\tilde{g},\tilde{K})$, $\sigma$ and $\tilde{\Ef}$ instead of the variables 
$(g,K)$, $k$ and $\Ef$.}
\end{Remark}

\n {\bf Proof:}

The crucial fact is to note that there are $\delta(R,\|Ric\|_{L^{2}_{g}(\Sigma)})$ and $\epsilon(R,\|Ric\|_{L^{2}_{g}(\Sigma)})$ such that if 
$\|g(k)-g_{0}\|_{H^{2}_{g_{0}}(B_{g_{0}}(p,R))}\leq \epsilon$ then $r_{2,g_{k}}(\partial B_{g_{0}}(p,R))\geq \delta$. This result can be easily proved by contradiction or simply invoking the discussion in \cite{A4} (see pgs. 218 and 227). Recall that
\ben
\|Ric\|^{2}_{L^{2}_{g}(\Sigma)}\leq C(|k|\Nu+Q_{0}),
\een

\n where $C$ is a numeric constant. As a result the $H^{2}$-harmonic radius of the region $B_{g(k)}(B_{g_{0}}(p,R),\frac{2}{3}\delta)$ is controlled
from below by $\La,\Nu_{0}$ and $R$ as long as $\|g(k)-g_{0}\|_{H^{2}_{g_{0}}(B_{g_{0}}(p,R))}\leq \epsilon$. Therefore \cite{Rei3} the norms $\|\hat{K}\|_{H^{2}_{g(k)}(B_{g_{0}}(p,R))}$, $\|N\|_{H^{3}_{g(k)}(B_{g_{0}}(p,R))}$, $\|E_{0}\|_{H^{1}_{g(k)}(B_{g_{0}}(p,R))}$ and $\|B_{1}\|_{H^{1}_{g(k)}(B_{g_{0}}(p,R))}$ are controlled from above by $\La$, $\Nu_{0}$ and $R$ as long as $\|g(k)-g_{0}\|_{H^{2}_{g_{0}}(B_{g_{0}}(p,R))}\leq \epsilon$. Under zero shift, the time derivatives of $g$ and $K$ are
\ben
g\dod=-2NK,
\een
\ben
K\dod=-\nabla^{2}N+N(E-K\circ K).
\een

\n Thus $\|g\dod\|_{H^{2}_{g(k)}(B_{g_{0}}(p,R))}$ and $\|K\dod\|_{H^{1}_{g_{0}}(B_{g_{0}}(p,R))}$ are controlled above by (say) $\tilde{\La}(\La,\Nu_{0},R)$ as long as 
$\|g(k)-g_{0}\|_{H^{2}_{g_{0}}(B_{g_{0}}(p,R))}\leq \epsilon$. Write
\ben
\|g(k)-g_{0}\|_{H^{2}_{g_{0}}(B_{g_{0}}(p,R))}\leq \int_{k_{0}}^{k}\|g\dod\|_{H^{2}_{g_{0}}(B_{g_{0}}(p,R))}dk,
\een
\ben
\|K(k)-K_{0}\|_{H^{1}_{g_{0}}(B_{g_{0}}(p,R))}\leq \int_{k_{0}}^{k}\|K\dod\|_{H^{1}_{g_{0}}(B_{g_{0}}(p,R))}dk.
\een

\n By Proposition \ref{ELEM} we can bound $\|g\dod\|_{H^{2}_{g_{0}}(B_{g_{0}}(p,R))}$ by $C_{1}\|g\dod\|_{H^{2}_{g(k)}(B_{g_{0}}(p,R))}$ and similarly for the $H^{1}_{g_{0}}$-norm of $K\dod$. Thus the length $\delta k$ of the maximal interval $[k_{0},k_{0}+\delta k]$ where $\|g(k)-g_{0}\|_{H^{2}_{g_{0}}(B_{g_{0}}(p,R))}\leq \epsilon$ is greater than $\epsilon/(C_{1}\tilde{\La})$ and similarly for the $H^{1}_{g_{0}}$-norm of $K$.\ep  

\begin{Prop}\label{KRIEBB}
Let $\Sigma$ be a compact three-manifold with $Y(\Sigma)\leq 0$. Assume $(\tilde{g},\tilde{K})$ is a cosmologically normalized long-time flow. Assume too that $\tilde{\Ef}\leq \La$ with $\La$ a positive constant. Then, for every $\epsilon>0$ and $R>0$ there exists $\sigma_{0}$ such that for any $\sigma\geq \sigma_{0}$ and $p\in \Sigma$ with $r_{2,\g(\sigma)}\geq 4R$ we have
\beq\label{K0}
\|\hat{\K}(\sigma)\|_{H^{1}_{\g(\sigma)}(B_{\g(\sigma)}(p,R))}\leq \epsilon,
\eeq
\beq\label{Ri0}
\|\hat{Ric}(\sigma)\|_{L^{2}_{\g(\sigma)}(B_{\g(\sigma)}(p,R))}\leq \epsilon,
\eeq
\beq\label{EB0}
\|E_{0}(\sigma)\|^{2}_{L^{2}_{\g(\sigma)}(B_{\g(\sigma)}(p,R))} +\|B_{0}(\sigma)\|_{L^{2}_{\g(\sigma)}(B_{\g(\sigma)}(p,R))}\leq \epsilon.
\eeq

\end{Prop}
 
\n {\bf Proof:}

The way to prove Proposition \ref{KRIEBB} is to show that for any $R>0$, any sequence of points $\{p_{i}\}$, and any divergent sequences of logarithmic-times $\{\sigma_{i}\}$ for which $r_{2,\g(\sigma_{i})}\geq 4R$, the norms (\ref{K0}), (\ref{Ri0}) and (\ref{EB0}) (with $\sigma_{i}$ instead of $\sigma$ and $p_{i}$ instead of $p$) tend to zero. We will use the terminology ``{\it Case \ref{K0}}" for the proof of this on $\hat{\K}$ and similarly for
$\hat{Ric}$ ({\it Case \ref{Ri0}}) and $E_{0},\ B_{0}$ ({\it Case \ref{EB0}}).

Let us start by making some elementary but important observations.  
\begin{Obs} \label{Obs1}{\rm From (the proof of ) Proposition \ref{CF} we know that there are $\{\delta \sigma_{i}\}$
with $|\delta \sigma_{i}|$ controlled from below by $\La$, $\Nu$ and $R$ (observe that because $\Nu$ is monotonic along the flow we can replace the dependence on $\Nu(\sigma_{i})$ for the dependence only on $\Nu_{0}=\Nu(\sigma_{0})$ with $\sigma_{0}$ some initial logarithmic time) and such that the norms
$\|\hat{\K}\|_{H^{2}_{\g(\sigma)}(B_{\g(\sigma_{i})}(p_{i},2R))}$ for $\sigma\in[\sigma_{i},\sigma_{i}+\delta\sigma_{i}]$, are controlled from above by $\La$, $\Nu_{0}$ and $R$. It follows
from the maximum principle applied to the lapse equation 
\ben
-\Delta_{\g(\sigma)}\N+|\K(\sigma)|^{2}_{\g(\sigma)}\N=1,
\een

\n that $\N(p,\sigma)\geq \N_{0}(\La,\Nu_{0},R)>0$ for $p$ in $B_{\g(\sigma_{i})}(p_{i},\frac{7}{4}R)$ and for $\sigma$ in $[\sigma_{i},\sigma_{i}+\delta \sigma_{i}]$\footnote{The argument is by contradiction. Assume there exists a sequence of states violating the inequality an obtain a convergent sub-sequence which violated the maximum principle.}.} 
\end{Obs}

\begin{Obs}\label{Obs2} {\rm Recall that
\ben
\frac{d\Nu}{d\sigma}=3\int_{\Sigma}3\N-1 dv_{\g}=3\int_{\Sigma}\phi dv_{\g},
\een

\n where (as was introduced in the background) $\phi=3\N-1$ is the Newtonian potential and satisfies $-1\leq \phi\leq 0$. If we integrate this equation between $\sigma_{i}$ and $\sigma_{i}+\delta \sigma_{i}$ (where $\delta \sigma_{i}$ will be the one in Proposition \ref{CF}) we get
\ben
\begin{split}
\Nu(\sigma_{i})-\Nu(\sigma_{i}+\delta \sigma_{i})=&-3\int_{\sigma_{i}}^{\sigma_{i}+\delta\sigma_{i}}\int_{\Sigma}\phi(\sigma)dv_{\g(\sigma)}d\sigma\\
&\geq -3\int_{\sigma_{i}}^{\sigma_{i}+\delta \sigma_{i}}\int_{B_{\g(\sigma_{i})}(p_{i},2R)}\phi(\sigma)dv_{\g(\sigma)}d\sigma.
\end{split}
\een

\n As $\Nu(\Sigma_{i})-\Nu(\sigma_{i}+\delta \sigma_{i})\rightarrow 0$ when $\sigma_{i}\rightarrow \infty$ (because $\Nu$ is monotonic and greater than zero) it follows that 
\ben
\mu\{\sigma \in [\sigma_{i},\sigma_{i}+\delta\sigma_{i}]/(\int_{B_{\g(\sigma_{i})}(p_{i},2R)}\phi^{2}(\sigma)dv_{\g(\sigma)})>\Gamma\}\rightarrow 0,
\een

\n as $\sigma_{i}\rightarrow \infty$, and for any fixed $\Gamma>0$.}
\end{Obs}

Let us prove now that $\|\hat{\K}\|_{L^{2}_{\g(\sigma_{i})}(B_{\g(\sigma_{i})}(p_{i},\frac{7}{4}R))}\rightarrow 0$ as $\sigma_{i}\rightarrow \infty$. Recall that
\ben
\frac{d\Nu}{d\sigma}=-3\int_{\Sigma}\N|\hat{\K}|^{2}_{\g}dv_{\g}.
\een

\n Integrating in $\sigma$ we have
\ben
\Nu(\sigma_{i})-\Nu(\sigma_{i}+\delta\sigma_{i})\geq 3 \int_{\sigma_{i}}^{\sigma_{i}+\delta\sigma_{i}}\int_{B_{\g(\sigma_{i})}(p_{i},\frac{7}{4}R)}\N|\hat{\K}|^{2}_{\g}dv_{\g}.
\een

\n It follows from Proposition \ref{CF} and {\it Observation \ref{Obs1}} that $\Nu(\sigma)$ can get below its limit $\Nu_{\infty}=\lim_{\sigma\rightarrow \infty}\Nu(\sigma)$ unless
$\lim_{\sigma\rightarrow \infty} \|\hat{\K}\|_{L^{2}_{\g(\sigma_{i})}(B_{\g(\sigma_{i})}(p_{i},\frac{7}{4}R))}=0$. Now as 

\n $\|\hat{\K}\|_{H^{2}_{\g(\sigma_{i})}(B_{\g(\sigma_{i})}(p_{i},2R))}$ is 
controlled from above by $\La$, $\Nu$ and $R$, it follows that if $\|\hat{\K}\|_{H^{1}_{\g(\sigma_{i})}(B_{\g(\sigma_{i})}(p_{i},\frac{7}{4}R))}\geq M>0$ we can extract a sub-sequence of the pointed spaces $(B_{\g(\sigma_{i})}(p_{i},frac{7}{4}R),p_{i},\g(\sigma_{i}))$ converging to a limit space (strongly in $H^{2}$) 

\n $(B_{\g_{\infty}}(p_{\infty},\frac{7}{4}R),p_{\infty},\g_{\infty})$ where
$\hat{\K}$ is not converging to zero which is a contradiction. This finishes the {\it case (\ref{K0})}. 

we use now this result and {\it Observation \ref{Obs1}} to get an improved version of {\it Observation \ref{Obs1}}.

\begin{Obs}\label{Obs3}{\rm Local elliptic estimates applied to the lapse equation (in the $\phi$-variable)
\ben
\Delta_{\g}\phi-|\K|^{2}_{\g}\phi=|\hat{\K}|^{2}_{\g},
\een

\n give
\ben
\mu\{\sigma\in[\sigma_{i},\sigma_{i}+\delta\sigma_{i}]/\|\phi\|_{H^{2}_{\g(\sigma)}(B_{\g(\sigma_{i})}(p_{i},\frac{3}{2}R))}(\sigma)\geq \Gamma\}\rightarrow 0,
\een

\n as $\sigma_{i}\rightarrow \infty$ and for any fixed $\Gamma>0$. An important consequence of this is that for any space-like tensors $U_{k}$, $k=1,2,3$ such that $\|U_{k}\|_{L^{2}_{\g(\sigma)}(B_{\g(\sigma_{i})}(p_{i},\frac{3}{2}R))}\leq M$ for some $M>0$ and for any 
$k=1,2,3$ we have
\ben
|\int_{\sigma_{i}}^{\sigma_{i}+\delta\sigma_{i}}\int_{B_{\g(\sigma_{i})}(p_{i},\frac{3}{2}R)}U_{0}*\phi+U_{1}*\nabla \phi+U_{3}*\nabla^{2}\phi \ dv_{\g(\sigma)}d\sigma|\rightarrow 0,
\een

\n as $\sigma_{i}\rightarrow \infty$.}
\end{Obs}

Recalling that
\ben
Curl_{\g} \Kt=-B_{0},
\een

\n we conclude that $\|B_{0}\|_{L^{2}_{\g(\sigma_{i})}(B_{\g(\sigma_{i})}(p_{i},2R))}\rightarrow 0$ (which is ``half" the {\it case (\ref{EB0})}). 

To prove the case (\ref{Ri0}) we note that it is enough from
\ben
\hat{Ric}_{\g}=E+\hat{\K}+\hat{\K}\circ \hat{\K}-\frac{1}{3}|\hat{\K}|^{2}\g,
\een

\n and {\it case (\ref{K0})}, to prove that $\|E\|_{L^{2}_{\g(\sigma_{i})}(B_{\g(\sigma_{i})}(p_{i},R))}$ tends to zero as $\sigma_{i}\rightarrow \infty$. This is however more difficult than the cases before. We will study the quantity
\ben
\int_{B_{\g(\sigma_{i})}(p_{i},\frac{3}{2}R)}<E,\hat{\K}>_{\g}dv_{g},
\een

\n and its time derivative with respect to logarithmic time. Differentiating with respect to 
$\sigma$ we have
\beq\label{inte}
\begin{split}
(\int_{B_{\g(\sigma_{i})}(p_{i},\frac{3}{2}R)}<E,\hat{\K}>_{\g}dv_{\g})\dod=&\int_{B_{\g(\sigma_{i})}(p_{i},\frac{3}{2}R)}<E\dod,\hat{\K}>_{\g}+<E,\hat{\K}>_{\g}\\
&-<E\circ \hat{\K},\g\dod>_{\g}+3<E,\hat{\K}>_{\g}\phi dv_{\g}.
\end{split}
\eeq

\n To get a more convenient expression of the right hand side of the previous equation we
will use the following expressions for the time derivatives of the cosmologically normalized variables $\g$, $E$ and $\hat{\K}$
\begin{equation}\label{gdot}
\dot{\g}=2\phi \g-6\tilde{N}\Kt,
\end{equation}
\begin{equation}\label{Edot}
\dot{E}=\N Curl_{\g} B -\frac{\nabla \N}{\N}\wedge_{\g} B-\frac{5}{2}E\times_{\g} \K-\frac{2}{3}<E,\K>_{\g}\g-\frac{3}{2}E,
\end{equation}
\begin{equation}\label{Kdot}
\dot{\Kt}=-\Kt-\phi\g-\nabla^{2}\phi+\phi E+E-\N(\Kt\circ\Kt-2\Kt).
\end{equation}

\n We now integrate equation (\ref{inte}) in $\sigma$ for $\sigma$ in $[\sigma_{i},\sigma_{i}]$. After integration of the left hand side we have (naturally) the expression 
\beq\label{lefths}
(\int_{B_{\g(\sigma_{i})}(p_{i},\frac{3}{2}R)} <E,\hat{\K}>_{\g}dv_{\g})(\sigma_{i}+\delta \sigma_{i})-(\int_{B_{\g(\sigma_{i})}(p_{i},\frac{3}{2}R)}<E,\hat{\K}>_{\g}dv_{\g})(\sigma_{i}).
\eeq

\n From {\it Case (\ref{K0})} and the bound
\ben
|\int_{B_{\g(\sigma_{i})}(p_{i},\frac{3}{2}R)}<E,\hat{\K}>_{\g}dv_{\g}|(\sigma)\leq 
\|E_{0}\|_{L^{2}_{\g}(B_{\g(\sigma_{i})}(p_{i},\frac{3}{2}R))}(\sigma)\|\hat{\K}\|_{L^{2}_{\g}(B_{\g(\sigma_{i})}(p_{i},\frac{3}{2}R))}(\sigma),
\een

\n we get that for any $\sigma$ (in particular for $\sigma=\sigma_{i}$ and $\sigma=\sigma_{i}+\delta \sigma_{i}$) we have that (\ref{lefths}) tends to zero as $i\rightarrow \infty$. Similarly, using either {\it Observation \ref{Obs3}}, {\it Case (\ref{K0})} or the $B_{0}$-part of {\it Case (\ref{EB0})} we have that all the terms in the right hand side of the integral in $\sigma$ of equation (\ref{inte}), except perhaps the term
\ben
\int_{\sigma_{i}}^{\sigma_{i}+\delta\sigma_{i}}\int_{B_{\g(\sigma_{i})}(p_{i},\frac{3}{2}R)}|E_{0}|^{2}dv_{\g}d\sigma,
\een

\n tend to zero. Thus we are lead to conclude that this term also tends to zero when $i\rightarrow \infty$. We will see now using the Gauss equation that $(\int_{B_{\g(\sigma_{i})}(p_{i},R)}|E_{0}|^{2}dv_{\g})(\sigma_{i})$ tends to zero as $i\rightarrow \infty$. That would finish {\it Case (\ref{EB0})} and {\it Case (\ref{Ri0})}. The argument is as follows. Consider a fixed, even and positive function $f$ of one variable $x$, equal to zero for $|x|\geq \frac{3}{2}$ and equal to one for $|x|\leq 1$. Consider the function $f(r)$ where $r$ is the geodesic radius from $p_{i}$ and corresponding to the metric $\g(\sigma_{i})$ inside $B_{\g(\sigma_{i})}(p_{i},2R)$. Extend $f(r)$ to the space-time in such a way that it is time independent. Consider finally the Weyl field $\W=f{\bf Rm}$. We have
\ben
E_{W}=fE_{0},\ B_{W}=fB_{0},
\een

\n and 
\ben
{\bf J}_{\W,bcd}=(\bn^{a}f){\bf Rm}_{abcd}.
\een

\n Thus, the $L^{2}_{\g}(B_{\g(\sigma_{i})}(p_{i},\frac{3}{2}R))$ norm of $E_{\W}$, $B_{\W}$ and ${\bf J}_{\W}$ are controlled by $\La$, $\Nu_{0}$ and $R$. It follows from integrating the
Gauss equation 
\ben
\tilde{Q}(\W)\dod=\tilde{Q}(\W)-9\int_{\Sigma}\N \tilde{Q}(\W)_{\alpha\beta \Ti\Ti}\tilde{\bf \Pi}^{\alpha\beta}dv_{\g}.
\een

\n in $\sigma$ and from $\sigma_{i}$ to $\sigma_{i}+\delta \sigma$ that
\ben
|\tilde{Q}(\W)(\sigma_{i}+\delta\sigma)-\tilde{Q}(\W)(\sigma_{i})|\leq \tilde{\La}(\La,\Nu_{0},R)\delta\sigma.
\een

\n Thus if $(\int_{B_{\g(\sigma_{i})}(p_{i},R)}|E_{0}|^{2}_{\g}dv_{\g})(\sigma_{i})\geq M>0$
we can choose $\delta\sigma$ such that for all $\sigma$ in $[\sigma_{i},\sigma_{i}+\delta\sigma]$ it is $\tilde{Q}(\W)(\sigma)\geq \frac{M}{2}>0$. But we have 
\ben
\tilde{Q}(\W)=\int_{B_{\g(\sigma_{i})}(p_{i},\frac{3}{2}R)}f^{2}(|E_{0}|^{2}+|B_{0}|^{2})dv_{\g},
\een

\n and we know from the $B_{0}$-part of {\it Case (\ref{EB0})} that
\ben
\lim_{\sigma_{i}\rightarrow \infty}\sup_{\sigma \in [\sigma_{i},\sigma_{i}+\delta\sigma]}\{(\int_{B_{\g(\sigma_{i})}(p_{i},\frac{3}{2}R)}f^{2}|B_{0}|^{2}dv_{\g})(\sigma)\}\rightarrow 0,
\een

\n when $i\rightarrow \infty$. Therefore, if $\sigma_{i}$ is big enough 
\ben
(\int_{B_{\g(\sigma_{i})}(p_{i},\frac{3}{2}R)}|E_{0}|^{2}dv_{\g})(\sigma)\geq \frac{M}{3}
\een

\n for any $\sigma$ in $[\sigma_{i},\sigma_{i}+\delta\sigma]$ which would contradict that
\ben
\int_{\sigma_{i}}^{\sigma_{i}+\delta\sigma_{i}}\int_{B_{\g(\sigma_{i})}(p_{i},\frac{3}{2}R)}|E_{0}|^{2}dv_{\g}d\sigma,
\een

\n tends to zero as $\sigma_{i}$ tends to infinity.\ep

We are ready to prove Theorem \ref{LTGFT}. The proof goes essentially along the same lines as the proof of the geometrization of the flow given in \cite{Rei2} for long time flows under $C^{\alpha}_{\g}$ curvature bounds. We repeat it here for the sake of clarity.

\vspace{0.2cm}
\n {\bf Proof (of Theorem \ref{LTGFT}):}

We prove first there is a divergence sequence of logarithmic times $\{\sigma_{i}\}$ with 

\n $(\Sigma^{\frac{1}{i}},(\g,\K)(\sigma_{i}))$ converging to $\cup_{i=1}^{i=n}(H_{i},(\g_{H,i},-\g_{H,i}))$ (weakly in $H^{2}$). Introduce a new variable $j=1,2,3,\ldots$. For $j=1$ find a sequence $\{\sigma_{1,i}\}$ with 
$(\Sigma^{1},\g(\sigma_{1,i}))$ convergent weakly in $H^{2}$. For $j=2$ find a sub-sequence $\{\sigma_{2,i}\}$ 
of $\{\sigma_{1,i}\}$ with $(\Sigma^{1/2},\g(\sigma_{2,i}))$ convergent in the weak $H^{2}$ topology. Proceed similarly
for all $j$ to have a double sequence $\{\sigma_{j,i}\}$. Now, for the diagonal sequence $\{\sigma_{i,i}\}$, $(\Sigma^{1/i},\g(\sigma_{i,i}))$
converges into a union of Riemannian manifolds of finite volume, denoted as $\cup_{\nu} (M_{\nu},\g_{\infty,\nu})$. 
By Proposition \ref{KRIEBB}, $\K(\sigma_{i,i})$ converges strongly to $-\g_{\infty,\nu}$ in $H^{1}$. Also by Proposition \ref{KRIEBB} we get that each metric $\g_{\infty,\nu}$ is hyperbolic and the convergence is in the strong $H^{2}$-topology. Therefore, as there
is a lower bound for the volume of complete hyperbolic manifolds of finite volume and the total volume of the limit 
space is bounded above, there must be a finite number of components, and we can write $\cup_{\nu} (M_{\nu},\g_{\infty,\nu})=\cup_{i=1}^{i=n}(H_{i},\g_{H,i})$.
 
We prove next that each component $(H_{j},\g_{H,j})$ is persistent. For simplicity assume there is only one component
and therefore $(\Sigma^{1/i},\g(\sigma_{i,i}))$ converges in the strong $H^{2}$-topology to $(H,\g_{H})$. There are two possibilities according to whether
the component is compact or not, we discuss them separately.

1.{\it (The compact case)} Assume $(H,\g_{H})$ is compact. Consider the space of metrics ${\mathcal{M}}_{H}$ in $H$. For every metric $g$ consider
the orbit of $g$ under the diffeomorphism group (of $H^{3}$-diffemorphisms). Denote such orbit by $o(g)$. Around $\g_{H}$
consider a small (smooth) section ${\mathcal{S}}$ of ${\mathcal{M}}_{H}$ (made of $H^{2}_{\g_{H}}$ metrics) and transversal to the 
orbits generated by the action on ${\mathcal{M}}_{H}$ of the diffeomorphism group
\footnote{Which particular section is taken is unimportant. One can use for instance ${\mathcal{S}}=\{g/id:(H,g)\rightarrow (H,\g_{H})\}$
is harmonic (see \cite{AM}, \cite{Ham}).}. If $\epsilon_{0}$ is sufficiently small 
every metric $g$ in ${\mathcal{M}}_{H}$ with $\|g-\g_{H}\|_{H^{2}_{\g_{H}}}\leq \epsilon_{0}$ can be uniquely projected 
into ${\mathcal{S}}$
by a diffeomorphism, or in other words we can consider the projection $P(g)=o(g)\cap {\mathcal{S}}$. Note that one can project
every flow of metrics $\g(t)$ starting close to $\g_{H}$, to a path $P(\g(t))$, until at least the first time when $\|P(\g(t))-\g_{H}\|_{H^{2}_{\g_{H}}}= \epsilon_{0}$
or in other words until at least when the projection touches the boundary of the ball of center $\g_{H}$ and radius $\epsilon_{0}$ in $H^{2}_{\g_{H}}$ (denote such ball as
$B(\g_{H},\epsilon_{0})$).

Recall Mostow rigidity\footnote{Mostow rigidity says that any two hyperbolic metrics on a compact manifold are necessarily isometric. What we 
state as {\it Mostow rigidity} here is an obvious consequence of this fact.}

\vspace{0.1cm}
\n {\it Mostow rigidity (the compact case)}. There is $\epsilon_{1}$ such that if $P(g'_{H})\in B(\g_{H},\epsilon_{1})$, where
$g'_{H}$ is a hyperbolic metric in $H$ then $P(g'_{H})=\g_{H}$. 

\vspace{0.1cm}
Fix $\epsilon_{2}=min\{\epsilon_{0},\epsilon_{1}\}$. Observe that as $\g_{\sigma_{i,i}}\rightarrow \g_{H}$ in $H^{2}$ there is 
a sequence of diffeomorphisms $\phi_{i}$ such that $\phi^{*}_{i}(g(\sigma_{i,i}))$ converges to $\g_{H}$ in $H^{2}_{\g_{H}}$.
Now, if the geometrization is not persistent there is $\epsilon\leq \epsilon_{2}$ and $i_{2}$ such that if $i\geq i_{2}$ then 
$P(\phi^{*}_{i}(\g(\sigma)))$ is well defined for $\sigma\geq \sigma_{i,i}$ until a first time $\sigma_{i,i}+T_{i}$ when 
$P(\phi^{*}_{i}(\g(\sigma_{i,i}+T_{i})))$ is in $\partial B(\g_{H},\epsilon_{2})$. But we know the sequence of Riemannian
manifolds $(H,P(\phi^{*}_{i}(\g(\sigma_{i,i}+T_{i}))))$ converge in $H^{2}$ to $\g_{H}$, and that means by the definition of $H^{2}$ convergence and 
Mostow rigidity that there is a sequence of diffeomorphisms
$\varphi_{i}$ such that $P(\varphi^{*}_{i}(P(\phi^{*}_{i}(\g(\sigma_{i,i}+T_{i}))))$ converges to $\g_{H}$ in 
$H^{2}_{\g_{H}}$. This contradict the fact that $P(\phi^{*}_{i}(\g(\sigma_{i,i}+T_{i})))$ is in $\partial B(\g_{H},\epsilon_{2})$.

2. {\it (The non-compact case)}. The proof 
of this case proceeds along the same lines as in the compact case but special care must be taken at the cusps\footnote{In \cite{Rei2} we have used CMC tori (transversal to the cusp) of a given area to to compare (in a unique way) the Riemannian spaces $(H,\g_{H})$ and $(\Sigma,\g(\sigma))$ (see \cite{Rei2} for details). If $\g(\sigma)$ is close to $\g_{H}$ only in $H^{2}_{\g_{H}}$ the CMC tori of a given area and transversal to the tori may be difficult to guarantee. It is for this reason that (see later in the text) we smooth out the metrics $\g(\sigma)$ nearby the regions where ``the CMC tori of a given area would be".}. Let us assume for 
simplicity that there is only one cusp in the piece $(H,\g_{H})$. Given $A$
sufficiently small there is a unique torus transversal to the cusp, to be denoted by $T^{2}_{A}$, of constant mean curvature 
and area $A$. Denote by $H_{A}$ the ``bulk" side of the torus $T^{2}_{A}$ in $H$. Consider the set ${\mathcal{M}}_{H_{A}}$ of metrics $\g$ on $H_{A}$ such that $\g=\g_{H}$ on $B_{\g_{H}}(T^{2}_{A},1)$. Consider the action of the diffeomorphism
group (of $H^{3}$-diffeomorphisms) on ${\mathcal{M}}_{H_{A}}$ and leaving 
$B_{\g_{H}}(T^{2}_{A},1)$ invariant. Again the orbit of a metric $\g$ will be denoted by $o(\g)$. Consider a small (smooth) section ${\mathcal{S}}$ of ${\mathcal{M}}_{H_{A}}$ transversal to the orbits of the action by the diffeomorphism group mentioned above. Finally consider the projection $P(\g)=o(\g)\cap {\mathcal{S}}$ which is
well defined on a ball $B(\g_{H},\epsilon_{0})$ for $\epsilon_{0}$ small enough. Observe again that a flow of metrics $\g(t)$ in ${\mathcal{M}}_{H_{A}}$ can be projected into ${\mathcal{S}}$ until at least the first time 
when $P(\g(t))$ is in $\partial B_{\g_{H}}(\g_{H},\epsilon_{0})$. Slightly abusing the notation (as we would require a pointed sequence) consider the sequence $(\Sigma,\g(\sigma_{i,i}))$ converging in $H^{2}$ to $\g_{H}$. There is a sequence of diffeomorphisms (onto the image) 
$\phi_{\sigma_{i,i}}:H_{A}\rightarrow \Sigma$ such that $\|\phi^{*}_{\sigma_{i,i}}
(\g(\sigma_{i,i}))-\g_{H}\|_{H^{2}_{\g_{H}}}$ converges to zero. Note that if we have a map
$\phi_{\sigma}:H_{A}\rightarrow \Sigma$ such that $\|\phi_{\sigma}^{*}\g(\sigma)-\g_{H}\|_{H^{2}_{\g_{H}}}\leq 2\epsilon$ for $\epsilon$ sufficiently small then we can deform $\phi^{*}_{\sigma}\g(\sigma)$ to a metric $S(\phi^{*}_{\sigma}\g(\sigma))$ in ${\mathcal{M}}_{H_{A}}$ in such a way that (a) $S(\phi^{*}_{\sigma}\g(\sigma))=\phi^{*}_{\sigma}\g(\sigma)$ on $H_{e^{4}A}$, (b) inside $(H_{e^{2}A}-H_{e^{4}A})$ the metric $S(\phi^{*}_{\sigma}\g(\sigma))$ is chosen to minimize the $L^{2}_{S(\phi^{*}_{\sigma}\g(\sigma))}$-norm of the traceless part of its Ricci tensor. Note the following elementary fact: if $\epsilon$ is chosen small enough and we have a diffemorphism $\phi:(H_{A},\g_{H})\rightarrow (\Sigma,\g)$ with $\|\phi^{*}_{\sigma}\g-\g_{H}\|_{H^{2}_{\g_{H}}}\leq 2\epsilon$ and $\phi^{*}\g$ is isometric to $\g_{H}$ then the new metric $S(\phi^{*}\g)$ is the deformation of $\g_{H}$ by a diffemorphism on $H_{A}$ (we will recall this note later as {\it note N}).

We make now the following crucial facts (justified below).

i. For all $A>0$ but sufficiently small there exists $\sigma_{0}$ and $\epsilon_{0}$ such
that for all $\epsilon\leq \epsilon_{0}$ and $\sigma_{1}\geq \sigma_{0}$ if there exists 
$\phi_{\sigma_{1}}:H_{e^{4}A}\rightarrow \Sigma$ with $\|\phi^{*}_{\sigma_{1}}\g(\sigma_{1})-\g_{H}\|_{H^{2}_{\g_{H}}}\leq \epsilon$ then there exists $\bar{\phi}_{\sigma_{1}}:H_{A}\rightarrow \Sigma$ with $\|\bar{\phi}^{*}_{\sigma_{1}}\g(\sigma_{1})-\g_{H}\|_{H^{2}_{\g_{H}}}\leq 2\epsilon$. Note that in this case $S(\bar{\phi}^{*}_{\sigma}\g(\sigma))$ is well defined.

ii. From i. we conclude that if we have $\phi_{\sigma_{1}}:H_{A}\rightarrow \Sigma$ with  satisfying: (a) $\|\phi^{*}_{\sigma_{1}}\g(\sigma_{1})-\g_{H}\|_{H^{2}_{\g_{H}}}\leq 2\epsilon$, (b) the restriction of $\phi_{\sigma_{1}}$ to $H_{e^{4}A}$ with $\|\phi^{*}_{\sigma_{1}}\g(\sigma_{1})-\g_{H}\|_{H^{2}_{\g_{H}}(H_{e^{4}A})}\leq \epsilon$, (c) $\|P(S(\phi^{*}_{\sigma_{1}}\g(\sigma_{1})))-\g_{H}\|_{H^{2}_{\g}}\leq \epsilon$ then $\phi_{\sigma}:H_{A}\rightarrow \Sigma$ with the properties (a) and (b) exist for $\sigma\geq \sigma_{1}$ (and varying continuously) until at least the first time $\sigma_{2}$ for which
$\|P(S(\phi^{*}_{\sigma_{2}}\g(\sigma_{2})))-g_{H}\|_{H^{2}_{\g_{H}}}=\epsilon$.

Let us justify now claim i. Recall Mostow Rigidity.

\vspace{0.1cm}
\n {\it Mostow rigidity (the non-compact case)}. There is $A_{0}$ such that for any $A\leq A_{0}$ there is $\epsilon_{0}$ such
that if $(\Sigma',g'_{H})$ is a complete hyperbolic manifold of finite volume and $\phi:H_{A}\rightarrow \Sigma'$ is a
diffeomorphism onto the image satisfying $\|\phi^{*}(g_{H}')-\g_{H}\|_{H^{2}_{\g_{H}}}\leq \epsilon_{0}$ then $(\Sigma',g'_{H})$ is isometric to $(H,\g_{H})$\footnote{The justification of this claim is as follows. According to the Mostow-Prasad rigidity $g'$ and $\g_{H}$ will be isometric if we can prove that $\Sigma'$ is diffemorphic to $H$. If $\epsilon$ is chosen small enough this is equivalent to show that the number of cusps of $\Sigma$ and $\Sigma'$ are the same. This follows from the Margulis lemma and the fact that if $\|\phi^{*}g'-\g_{H}\|_{H^{2}_{\g_{H}}}\leq \epsilon$ then $\|\phi^{*}g'-\g_{H}\|_{C^{\frac{1}{2}}_{\g_{H}}}\leq C\epsilon$ where $C$ is a numeric constant.} 

\vspace{0.1cm}

The justification of i. follows straight from Mostow rigidity. Indeed pick any $A$ such that $e^{2}A\leq A_{0}$ and $\epsilon\leq \epsilon_{0}$ as in the Mostow rigidity statement. Suppose there exists a divergent sequence $\{\sigma_{i}\}$ and a sequence of diffeomorphisms onto the image 
$\phi_{\sigma_{i}}:H_{e^{4}A}\rightarrow \Sigma$ such that $\|\phi^{*}_{\sigma_{i}}\g(\sigma_{i})-\g_{H}\|_{H^{2}_{\g_{H}}}\leq \epsilon$ but such that it cannot be extended to a diffemorphism
$\bar{\phi}_{\sigma_{i}}:H_{A}\rightarrow \Sigma$ with $\|\bar{\phi}^{*}_{\sigma_{i}}-\g_{H}\|_{H^{2}_{\g_{H}}}\leq 2\epsilon$. We can extract a (pointed) sub-sequence of $\{(\Sigma,\g(\sigma_{i})\}$ converging to a complete hyperbolic metric of finite volume, which, by Mostow rigidity and the choice of $A$ and $\epsilon$ must be isometric to $\g_{H}$.
Therefore for $\sigma_{i}$ sufficiently big the diffeomorphism $\bar{\phi}_{\sigma_{i}}$ can be defined which is a contradiction. 

Now from facts i. and ii. we get that, if the 
geometrization $(H,\g_{H})$ is not persistent there is $\epsilon\leq \epsilon_{0}$ and $\sigma_{0}$ such that if $\sigma_{i,i}\geq \sigma_{0}$ then 
$P(S\phi^{*}_{\sigma}(\g(\sigma)))$ is well defined for $\sigma\geq \sigma_{i,i}$ until a first time $\sigma_{i,i}+T_{i}$ when 
$P(S(\phi^{*}_{i}\g(\sigma_{i,i}+T_{i})))$ is in $\partial B(\g_{H},\epsilon)$. Now the sequence $\phi^{*}_{i}(\g(\sigma_{i,i}+T_{i}))$ 
has a sub-sequence converging in $H^{2}$ to a complete hyperbolic metric of finite volume. Again as in the compact case, by Mostow rigidity it must be
converging in $H^{2}$ to $\g_{H}$. Therefore (recall {\it note N}) $P(S(\phi^{*}_{\sigma}\g(\sigma)))$ must be converging to a metric on $H_{A}$ which is a diffeomorphism of
$\g_{H}$ contradicting the fact that $P(S(\phi^{*}_{i}(\g(\sigma_{i,i}+T_{i}))))$ is in $\partial B_{\g_{H}}(\g_{H},\epsilon_{2})$.
 
To finish the proof of the persistence of the geometrization one still needs to show that the compliment of the persistent 
pieces $(H_{i},\g_{H,i})$ is the $G$ sector or in other words that for any $\epsilon>0$, $(\Sigma^{\epsilon}(\sigma),\g(\sigma))$ converges to 
the $\epsilon$-thick part of the persistent pieces 
$(H_{i},\g_{H,i})$. The proof of this fact follows by contradiction. If this is not the case one can extract a divergent sequence of logarithmic times containing 
an $H$-piece different  from the pieces $(H_{i},\g_{H,i})$. One can prove again that this new piece is persistent leading into 
a contradiction for if persistent, the piece must be one of the pieces $(H_{i},\g_{H,i})$  by the way these pieces 
are defined.  
\ep

{\center \subsection{Stability of the flat cone ({\it Case $Y(\Sigma)<0$ (I)}-ground state)}
\label{SFC}}

In this section we will prove the stability of the {\it Case $Y(\Sigma)<0$ (I)}-ground state. Namely, we will show that a cosmologically normalized flow $(\tilde{g},\tilde{K})$ 
over a hyperbolic three-manifold $\Sigma$, with initial data $(\tilde{g},\tilde{K})(\sigma_{0})$ close (in $H^{3}\times H^{2}$) to the ground state $(g_{H},-g_{H})$, converges (in $H^{3}\times H^{2}$) to the ground state $(g_{H},-g_{H})$ when $\sigma\rightarrow \infty$. This stability has been proved by Andersson and Moncrief in \cite{AM} (for rigid hyperbolic manifolds $\Sigma$\footnote{The rigidity condition is a somehow mild restriction. We remove it with an appropriate use of the reduced volume. The 
core of the proof is essentially the same as in \cite{AM}.}). 

\begin{T} {\rm (Stability of the flat cone).}\label{sfc} Let $\Sigma$ be a compact hyperbolic three-manifold. Then, there is an $\epsilon>0$ such that the cosmologically normalized CMC flow $(\tilde{g},\tilde{K})(\sigma)$ 
of a cosmologically normalized ($H^{3}\times H^{2}$) initial state $(g_{0},K_{0})=(\tilde{g}(\sigma_{0}),\tilde{K}(\sigma_{0}))$ with $\tilde{\Ef}(\sigma_{0})+({\mathcal{V}}-{\mathcal{V}}_{inf})\leq \epsilon$, converges in $H^{3}_{g_{H}}\times H^{2}_{g_{H}}$ (and for a suitable choice of the shift vector $X$) to $(g_{H},-g_{H})$ (the standard {\it Case $Y(\Sigma)<0$ (I)}-ground state).
\end{T}

\begin{Remark}{\rm As it is stated Theorem \ref{sfc} gives 
few information about the shift vector $X$. This inconvenient can be remedied if, as in \cite{AM}, $X$ is chosen in such a way that for every $\sigma$ the identity $id:(\Sigma,\g(\sigma))\rightarrow (\Sigma,g_{H})$ is a harmonic map (the {\it spatially harmonic gauge} \cite{AM}). Full control of the evolution of the shift vector $X$ can be obtained in this case.}
\end{Remark} 

We begin with a preliminary Proposition.

\begin{Prop}\label{rvc} Say $\Sigma$ is a compact hyperbolic three-manifold. Fix $\underline{\nu}_{0}>0$ and $\Nu_{0}>\Nu_{inf}$. Then, for every $\epsilon>0$ there is $\delta(\epsilon,\underline{\nu}_{0},\Nu_{0})>0$ such that for every cosmologically normalized state $(\g,\K)$ with $\underline{\nu}\geq \underline{\nu}_{0}$, $\Nu\leq \Nu_{0}$ and $\|\hat{K}\|_{L^{2}_{\g}}+\tilde{Q}_{0}\leq \delta$ we have ${\mathcal{V}}-{\mathcal{V}}_{inf}\leq \epsilon$.
\end{Prop}

\n {\bf Proof:} 

It is enough to prove that any sequence $(\g,\K)$ (we will forget about putting sub-index) with 
$\|\hat{\K}\|_{L^{2}_{\g}}+\tilde{Q}_{0}\rightarrow 0$ has a 
sub-sequence converging in $H^{2}$ to $g_{H}$. From \cite{Rei3} Proposition 3 we have (for arbitrary states $(g,K)$)
\ben
(\int_{M}2|\nabla\hat{K}|^{2}+|\hat{K}|^{4}dv_{g})^{\frac{1}{2}}\leq C(|k|\|\hat{K}\|_{L^{2}_{g}}+Q_{0}^{\frac{1}{2}}).
\een

\n Thus $\|\hat{\K}\|_{L^{4}_{\g}}\rightarrow 0$ as $\|\hat{\K}\|_{L^{2}_{\g}}+\tilde{Q}_{0}\rightarrow 0$. From this and
\ben
\hat{Ric}_{\g}=E+\hat{\K}+\hat{\K}\circ \hat{\K}-\frac{1}{3}|\hat{\K}|^{2}\g,
\een

\n we get that $\|\hat{Ric}_{\g}\|_{L^{2}_{\g}}\rightarrow 0$. Moreover from the energy constraint  we get $\|R_{\g}+6\|_{L^{2}_{\g}}\rightarrow 0$. As $\Nu$ is bounded above and $\underline{\nu}$ bounded below, there is a sub-sequence of $(\g,\K)$ converging (in $H^{2}$) to $g_{H}$. Thus $\Nu\rightarrow \Nu_{inf}$.\ep

\noindent {\bf Proof} (of theorem \ref{sfc}): 

Recall from Theorem \ref{GST} and \cite{Rei3} that for any $\epsilon>0$ there is $\delta>0$ such that if for a cosmologically normalized state $(\g,\K)$ it is $\tilde{\Ef}+(\Nu-\Nu_{inf})\leq \delta$ then there is a diffeomorphism $\phi$ such that $(\phi^{*}(\g),\phi^{*}(\K))$ is $\epsilon$-close to $(g_{H},-g_{H})$ in $H^{3}_{g_{H}}\times H^{2}_{g_{H}}$. One can also find $\delta>0$ such that in addition the $L^{\infty}_{\g}$-norm of the deformation tensor $\dt_{ab}=\bn_{a}T_{b}$ (with respect to the CMC foliation) is less than $\epsilon$. It is direct to see \cite{AM} that this implies the following inequality for the evolution of $\tilde{\Ef}$
\begin{equation}\label{eq:BR1}
\partial_{\sigma}\tilde{\mathcal{E}}_{1}\leq -(2-C\tilde{\mathcal{E}}_{1}^{\frac{1}{2}})\tilde{\mathcal{E}}_{1}.
\end{equation}

\noindent Thus, from it, the monotonicity of the reduced volume and the continuity principle of Theorem 1 in \cite{Rei3} we conclude that
the flow is a long-time flow. Note that the argument is independent of the shift $X$. One may well take the zero shift $X=0$. Now, it is clear from equation (\ref{eq:BR1}), that $\tilde{\mathcal{E}}_{1}\rightarrow 0$ as the logarithmic time diverges. To show that (up to diffeomorphism) the flow $(\g,\K)$ converges (in $H^{3}_{g_{H}}\times H^{2}_{g_{H}}$) to $(g_{H},-g_{H})$ it remains to prove 
that ${\mathcal{V}}-{\mathcal{V}}_{inf}\rightarrow 0$. By Proposition \ref{rvc} if ${\mathcal{V}}(\sigma)-{\mathcal{V}}_{inf}\geq \Gamma>0$ for all $\sigma$ (observe 
that ${\mathcal{V}}$ is monotonically decreasing) it must be $\|\tilde{\hat{K}}\|_{L^{2}_{\g}}(\sigma)\geq M>0$ (for some $M>0$) for all $\sigma\geq \sigma_{1}$. If $\epsilon$ is chosen small enough it is must be $\|\tilde{N}(\sigma)-\frac{1}{3}\|_{L^{\infty}}<\frac{1}{6}$ for all $\sigma\geq \sigma_{0}$. The equation for the evolution of the reduced volume
\begin{equation}
\frac{d{\mathcal{V}}}{d\sigma}=-3\int_{\Sigma}\tilde{N}|\tilde{\hat{K}}|^{2}dv_{\tilde{g}},
\end{equation}

\n shows that if $\|\tilde{\hat{K}}\|_{L^{2}_{\g}}(\sigma)\geq M>0$ for $\sigma\geq \sigma_{1}$ then ${\mathcal{V}}-{\mathcal{V}}_{\inf}$ must go below zero after some time which is a contradiction.\ep

{\center \section{Hyperbolic rigidity, ground states and gravitational waves}\label{HRGSGW}}

There are several theoretical reasons to believe that the reduced volume $\Nu$ should decrease to its infimum $\Nu_{\inf}=(-\frac{1}{6}Y(\Sigma))^{\frac{3}{2}}$ at least for solutions in the family of long time solutions having a uniform bound on $\tilde{\Ef}$. It may be possible (see \cite{Rei2}) to prove this claim for long time solutions having uniform bounds on the $C^{\alpha}_{\g}$-norm
of (the electric and magnetic parts of the) space-time Riemann tensor. Proving the claim for solutions in the family of long-time solutions having a uniform bound in $\tilde{\Ef}$ could be a task of much greater difficulty. In this section we present various facts and arguments pointing to the validity of this claim.

According to Margulis, hyperbolic cusps are rigid in the following sense: if a complete hyperbolic metric $\g_{H}$ on a manifold $\field{R}\times T^{2}$ is close enough to a hyperbolic cusp metric $\g_{C}=dx^{2}+e^{2x}g_{T^{2}}$ over a domain $\Omega=[-a,\infty)\times T^{2}$ with $a$ positive and big enough, then $\g_{H}$ is isometric to $\g_{C}$. 

Consider the following spaces
\ben
DC=\{\g\ {\rm on}\ \field{R}\times T^{2}/R_{\g}\geq -6,\ \g\sim \g_{R}\ {\rm when}\ x\rightarrow \infty\ {\rm and} \g\sim \g_{L}\ {\rm when}\ x\rightarrow -\infty\},
\een  
\ben
SC=\{\g\in {\rm on}\ D\times S^{1}/ R_{\g}\geq -6,\ {\rm and}\ g\sim \g_{S,R}\ {\rm when}\ 
x\rightarrow \infty\},
\een

\begin{figure}[h]
\centering
\includegraphics[height=11cm,angle=-90]{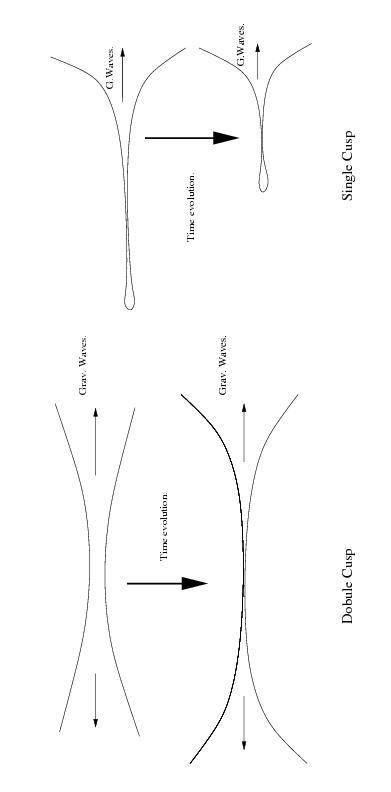}
\caption[DCSC]{The (conjectural) evolution of the Double Cusp and Single Cusp.}
\label{DCSC}
\end{figure}

\n where $DC$ accounts for ``double cusps" and $SC$ for ``single cusp". $\g_{R}$ and $\g_{L}$ are two arbitrary but fixed hyperbolic cusp metrics on the (right and left) ends of $\field{R}\times T^{2}$ and $\g_{S,R}$ is an arbitrary but fixed metric on the (right) end of $D\times S^{1}$ ($D$ is the unit two-dimensional disc). Consider now two cosmologically normalized flow $(\g_{DC},\K_{DC})$ and $(\g_{SC},\K_{SC})$ over $\field{R}\times T^{2}$ and $D\times S^{1}$ respectively and with $\g_{DC}$ in $DC$ and $\g_{SC}$ in $SC$ (see Figure \ref{DCSC}). As the states evolve one may argue that they lose ``energy" (actually they lose reduced volume) by the emission of cylindrical gravitational waves\footnote{In the definition of the sets $DC$ and $SC$ we can assume the metrics $\g$ are $T^{2}$-symmetric. That would justify the statement that the system emits cylindrical gravitational waves.} at the ends of the cusps. According to Margulis the states would settle into the infinite double cusp (for the flow $(\g_{DC},\K_{DC})$) or the infinite single cusp (for the flow $(\g_{SC},\K_{SC})$) if it were the case that these configurations are $\Nu$-rigid. This is indeed true for the double cusp (a ground state) but false for the single cusp 
(a non-ground state) in the following sense.

\vspace{0.2cm}
\begin{Prop}\label{DC}
Consider the set of metrics $\g$ in $DC$ with $\g=\g_{R}$ for $x\in [a_{R},\infty)$ and $\g=\g_{L}$ for
$x\in(-\infty,a_{L}]$. Call $\Nu_{R}$ the volume of $\g_{R}$ on the region $(-\infty,a_{R}]\times T^{2}$ and similarly for the left cusp ($\Nu_{L}$). Then
the volume of $\g$ on the region $[a_{L},a_{R}]\times T^{2}$ is strictly greater
than $\Nu_{L}+\Nu_{R}$.
\end{Prop}  

\begin{Prop}\label{SC}
Consider the set of metrics $\g$ in $SC$ with $\g=\g_{S,R}$ for $x\in [a_{R},\infty)$. Call $\Nu_{R}$ the volume of $\g_{S,R}$ on the region $(-\infty,a_{R}]\times T^{2}$. Then there exist metrics $\g$ as described above and having volume inside the region $(-\infty,a_{R}]\times T^{2}$ less than $\Nu_{R}$.
\end{Prop}

\n A proof of Proposition \ref{SC} and an explicit construction of such metrics is given in \cite{Rei4} (the metrics are indeed $T^{2}$-symmetric). It can be seen analytically (and numerically) that as time evolves the evolution of the (Yamabe) initial states $(\g_{0},-\g_{0})$ described in \cite{Rei4} actually separates from the single infinite hyperbolic cusp (as it should be).


\addcontentsline{toc}{section}{\bf Bibliography} 
\begin{thebibliography}{99}

\bibitem[1] {A4}Anderson, Michael T. Extrema of curvature functionals on the space of metrics on $3$-manifolds. 
{\em Calc. Var. Partial Differential Equations 5 (1997), no. 3, 199--269.}

\bibitem[2]{ASCS} Anderson, M. T. Scalar curvature, metric degenerations, and the static vacuum 
Einstein equations on 3-manifolds. II. {\em Geom. Funct. Anal. 11 (2001), no. 2, 273--381}.

\bibitem[3]{A3}Anderson, Michael T. Scalar curvature and the existence of geometric structures on 3-manifolds. I. {\em J. Reine Angew. Math. 553 (2002), 125--182}.

\bibitem[4]{AM} Andersson, Lars; Moncrief, Vincent. 
Future complete vacuum space-times. 
{\em The Einstein equations and the large scale behavior of gravitational fields, 299--330, 
BirkhÃ€user, Basel, 2004}.

\bibitem[5]{CHG} Cheeger, Jeff; Gromov, Mikhael Collapsing Riemannian manifolds while keeping their curvature bounded. I.  J. Differential Geom.  23  (1986),  no. 3, 309--346.

\bibitem[6]{CK} Christodoulou, Demetrios; Klainerman, Sergiu The global nonlinear stability of the Minkowski space. 
{\em Princeton Mathematical Series, 41. Princeton University Press, Princeton, NJ, 1993}.

\bibitem[7]{FM1} Fischer, Arthur E.; Moncrief, Vincent The reduced hamiltonian of general relativity and the 
$\sigma$-constant of conformal geometry. 
{\em Mathematical and quantum aspects of relativity and cosmology (Pythagoreon, 1998),
 70--101, Lecture Notes in Phys., 537, Springer, Berlin, 2000.} 

\bibitem[8]{GT} Gilbarg, David; Trudinger, Neil S. Elliptic partial differential equations 
of second order. {\it Reprint of the 1998 edition. Classics in
 Mathematics. Springer-Verlag, Berlin, 2001.}

\bibitem[9]{Ham} Hamilton, Richard S. Non-singular solutions of the Ricci flow on three-manifolds.  
{\it Comm. Anal. Geom.  7  (1999),  no. 4, 695--729.} in the book: Collected papers on Ricci flow.
Edited by H. D. Cao, B. Chow, S. C. Chu and S. T. Yau. Series in Geometry and Topology, 37. International 
Press, Somerville, MA, 2003.

\bibitem[10]{I} Isenberg, James, Constant mean curvature solutions of the Einstein 
constraint equations on closed manifolds. {\it Classical Quantum Gravity 12 (1995), no. 9, 
2249--2274.} 

\bibitem[11]{Rei4} Reiris, Martin. Dissertation Thesis: Aspects of the Long Time Evolution in General Relativity and Geometrizations of Three-Manifolds. {\it State University of New York at Stony Brook 2005.}

\bibitem[12]{Rei} Reiris, Martin General ${\mathcal{K}}=-1$ Friedman-Lemaître models and the averaging problem in cosmology. {\it Classical Quantum Gravity  25  (2008),  no. 8, 085001, 26 pp.}

\bibitem[13]{Rei2} Reiris, Martin. On the asymptotic spectrum of the reduced volume in cosmological solutions of the Einstein equations. {\it To appear in Gen. Rel. Grav.}

\bibitem[14]{Rei3} Reiris, Martin. The constant mean curvature Einstein flow and the Bel-Robinson energy. {\it Preprint}.

\bibitem[15]{Y1} Yang, Deane. Convergence of Riemannian manifolds with integral bounds on curvature. I.  {\it Ann. Sci. École Norm. Sup. (4)  25  (1992),  no. 1, 77--105.}

\bibitem[16]{Y2} Yang, Deane. Convergence of Riemannian manifolds with integral bounds on curvature. II.  {\it Ann. Sci. École Norm. Sup. (4)  25  (1992),  no. 2, 179--199}.

\end{thebibliography}
\end{document}